 \documentclass[twocolumn,12pt]{revtex4}
\usepackage{mathtools}
\usepackage{amsmath,bbm}
\usepackage{amssymb,bm}
\usepackage{array}
\usepackage{graphicx}
\usepackage{enumerate}

\usepackage{slashed}
\usepackage{caption}
\begin{document}
\title{Effect of temperature gradient on heavy quark anti-quark potential using gravity dual model}
\author{S. Ganesh\footnote{Corresponding author:\\Email: gans.phy@gmail.com}}
\author{M. Mishra}
\affiliation{Department of Physics, Birla Institute of Technology and Science, Pilani - 333031, INDIA}
\begin{abstract}
	%recipe

	%gansref_rev1
	Thermal systems have traditionally been modeled via Euclideanized space by analytical continuation of time to an imaginary time. We extend the concept to static thermal gradients by recasting the temperature variation as a variation in the Euclidean metric. We apply this prescription to determine the Quark anti-Quark potential in a system with thermal gradient. A naturally occurring QCD medium with thermal gradients is a Quark Gluon Plasma (QGP).
However, the QGP evolves in time. Hence, we use a quasi-stationary approximation, which is applicable only if the rate of time evolution is slow. The application of our proposal to a Quark anti-Quark potential in QGP can be seen as a step towards a more exact theory which would incorporate time varying thermal gradients.
	The effect of a static temperature gradient on the Quark anti-Quark potential is analyzed using a gravity dual model. 
A non-uniform black string metric is developed, by perturbing the Schwarzchild metric, which allows to incorporate the temperature gradient in the dual AdS space. 
Finally, an expression for the Quark anti-Quark potential, in the presence of a static temperature gradient, is derived.

\vskip 0.5cm

\end{abstract}
\maketitle
\section{Introduction}
\label{sec:intro}
	Much of fundamental physics has been developed for systems in vacuum, i.e. zero temperature. However, systems with finite temperature including non-uniform temperature are all pervasive. 
	A Euclideanized space obtained by an analytical continuation of time to imaginary time, is one of the mathematical techniques used to model thermal systems. 

	In this paper, we prescribe an extension, in which static thermal gradients are modeled by recasting the variation in temperature as a variation in the Euclidean metric.
	We then attempt to apply this hypothesis to a Quark anti-Quark immersed in a QGP. The calculations are performed in the gravity dual domain.
	In the dual domain, the corresponding variation in the metric is obtained by perturbing a black string. We then calculate the Quark anti-Quark potential using the perturbed black string metric.

	The QGP is a deconfined state of quarks and gluons, formed at very high energies. At the high energy hadron colliders like Relativistic Heavy Ion Collider (RHIC) or Large Hadron Collider (LHC), it is believed to be produced and manifests itself as an expanding fireball. The QGP is supposed to be in a state of local thermodynamic equilibrium and reaches this equilibrated state at about 1 fm/c after the heavy ions collide at RHIC \cite{nature}. At LHC, the equilibrated state is supposed to reach even sooner (about 0.1 to 0.5 fm/c) due to higher energies involved.
	As the QGP evolves in time, the static gradient hypothesis is not directly applicable, and a more exact model would be required which takes care of time variation in thermal gradients. However, as a step towards determining the Quark anti-Quark potential, under conditions of a thermal gradient, we use a quasi-stationary approximation. If the Quark anti-Quark had been in a static plasma, with no time dependence, our prescription would have been more appropriate.
 Relativistic hydrodynamics is used to model the QGP after equilibration, and hydrodynamic calculations for a non-viscous fluid, are in good agreement with experimental data~\cite{nature, naturephy, PRCver}.
Since viscosity is related to the mean free path of the constituent particles, these results strongly indicate an equilibrated QGP medium. 
The equilibrated nature of the QGP is used to justify the quasi-stationarity approximation in appendix~\ref{sec:quasi}.
In the quasi-stationary approximation, the duration of evolution is divided into small time intervals. Within a small time interval, the system is assumed to be static, i.e. the fluid is stationary. The effect of hydrodynamical evolution 
%and radial (or elliptic) flow 
is introduced by varying the system parameters across different time intervals. In the dual domain, this is accomplished by keeping the metric constant within a given time interval, and varying it across different time intervals. 
We believe this approach would lead to a more appropriate Quark anti-Quark potential when compared with using a constant temperature.
%In a thermal system at local thermodynamic equilibrium, the mean free path is much smaller than the rate at which the temperature changes.
	
	The suppression of  $J/\psi$ and $\Upsilon$ are two prominent signatures to detect the presence of QGP in heavy-ion collision experiments, as well as to study its properties~\cite{mats, Chu, PKSMCA, PKSRAR, PKSAAD, PKSCMS1, PKSBAB, Madhu1, gans1}. 
In~\cite{gans2,gans3}, it has been shown that at high temperatures, the heavy Quark anti-Quark separation increases, which leads to an increase in formation time for both $J/\psi$ and $\Upsilon$. This in turn results in a drastic reduction in suppression due to the color screening mechanism~\cite{gans2, gans3}. 
Thus, accurate determination of heavy Quark anti-Quark potential is essential to determine the $J/\psi$ and $\Upsilon$ suppression precisely. 
We hope that our proposed hypothesis should help improve the accuracy of determination of Quark anti-Quark potential in a QGP. Our hypothesis, related to spatial thermal gradients, is thus a step towards an ultimate thermal model which incorporates both temporal and spatial thermal gradients.

   If $\Theta(z,t)$ is the temperature of the system,  at time $t$ and location $z$, then  $\Theta(z,t)$ can be decomposed into a time independent $\Theta_0(z)$ and time dependent $\Delta \Theta(z,t)$, where $\Theta_0(z)$ is the time averaged value of $\Theta(z,t)$ over a small time interval $\Delta t$, i.e.,
\begin{equation}
\label{eq:tempgrad_def}
\Theta(z,t) = \Theta_0(z) + \Delta \Theta(z,t).
\end{equation}
It is the static temperature gradient $\Theta_0(z)$, that we attempt to model in this paper. 
% gans R3:
Let $v_z$ be the velocity of the thermal medium (Eg. the radial velocity in QGP).  
Then, considering only $\Theta_0(z)$ can be expected to provide reasonable accuracy, only if $\Delta \Theta(z,t) \approx \frac{d\Theta(z,t)}{dt}\Delta t = \left ( \frac{\partial \Theta(z,t)}{\partial t} + v_z\frac{\partial \Theta(z,t)}{\partial z} \right )  \Delta t$ is small. 
This implies a slow time varying system with both $\frac{\partial \Theta(z,t)}{\partial t}$ and velocity term $v_z\frac{\partial \Theta(z,t)}{\partial z}$ being small and the time interval is small.
In Sec.~\ref{sec:black_string_perturb}, it will be seen how the condition of slow time varying $\Delta \Theta(z,t)$, naturally emerges while developing the metric in the dual domain. 
At this point, we try to quantify slow time variations. The particles constituting the thermal medium (say QGP), would be in a state of constant collisions. Let $\tau_m$ be the mean free time of the collisions. We take a time interval $\Delta t \gg \tau_m$. For such a $\Delta t$, the condition 
\begin{equation}
\label{eq:approxcond}
	\left ( \frac{\partial \Theta(z,t)}{\partial t} + v_z\frac{\partial \Theta(z,t)}{\partial z} \right ) \Delta t \ll \Theta_0(z)~\forall~z, 
\end{equation}
should be satisfied. 
Thus, our prescription would be applicable only for those QGP or other time varying thermal systems, where the condition in Eq.~\ref{eq:approxcond} is satisfied.

The gravity dual model in the form of AdS-CFT correspondence was first proposed by Maldacena in his seminal work~\cite{mald1, mald2}. 
AdS-CFT has been an area of current research for understanding QGP properties~\cite{kovtun,hotwind,tatsuo2}  and other areas~\cite{tatsuo1, koji, hyo}. 
It has been shown that the strong coupling in the framework of Quantum Chromodynamics (QCD), corresponds to a weak coupling in the gravity dual domain, and hence calculations can be more easily be performed in the gravity dual domain. 
With the increase in the separation between the Quark and the anti-Quark at higher temperatures, the coupling constant would become larger, and hence, the potential cannot be calculated accurately using perturbative QCD.
The gravity dual model provides an ideal mathematical framework to determine the effect of a temperature gradient on the heavy Quark anti-Quark potential, when the separation between them is large. However, the gravity model in $AdS_5$ space is dual to {\fontfamily{pzc}\selectfont N} $= 4$ supersymmetric Yang-Mills Lagrangian, and hence our calculation would be for a {\fontfamily{pzc}\selectfont N} $= 4$ supersymmetric Yang-Mills Lagrangian instead of the QCD Yang-Mills Lagrangian. 

	The organization of the rest of the paper is as follows. Section~\ref{sec:polyakov_dual} frames the problem in both the Polyakov loop domain and in the gravity dual domain. 
Section~\ref{sec:black_string} dwells into the determination of the metric due to the insertion of a non-uniform black string, which is required for creating a temperature gradient in the dual domain.  
The calculation of the string action in the dual gravity domain, and hence the Quark anti-Quark potential, in the presence of a temperature gradient, is treated in Sec.~\ref{sec:tempgrad}. 
The summary and conclusions are finally drawn in Sec.~\ref{sec:conclusion}.

%gans
\section{Wilson-Polyakov Loop correlator and Dual Domain Formulation}
\label{sec:polyakov_dual}
%gans
\subsection{Wilson-Polyakov Loops}
\label{sec:polyakov}
\begin{figure}[h!]
%gans
%\includegraphics[width = 60mm,height = 60mm]{single_wil_loop_dbeta.eps}
\includegraphics[width = 60mm,height = 60mm]{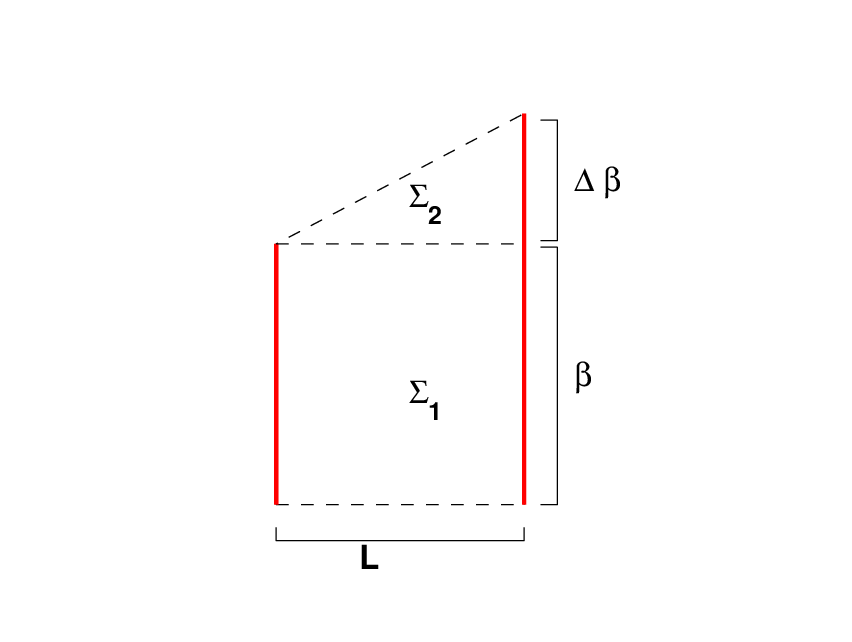}
\captionsetup{justification=raggedright, singlelinecheck=false}
%gans
%\caption{Wilson loop to model the potential of a heavy Quark anti-Quark pair immersed in a QGP medium with a temperature gradient.}
\caption{Polyakov loops to model the potential of a heavy Quark anti-Quark pair immersed in a thermal medium with a temperature gradient. 
%gans
The two solid vertical red lines represent the two Wilson lines corresponding to the two Polyakov loops.}
\label{fig:single_loop}
\end{figure}

The Wilson-Polyakov loops that has been used to model the scenario of the temperature gradient is depicted in Fig.~\ref{fig:single_loop}. The relation of the Wilson-Polyakov loop correlator to the heavy Quark anti-Quark potential could be modeled as~\cite{soo}:   
\begin{equation}
\label{eq:qqbeta}
%gans
%<W(C)> = e^{-V_{Q\bar{Q}}\beta},
<P(0)P(L)> = e^{-V_{Q\bar{Q}}\beta},
\end{equation}
where, $\beta = \frac{1}{\Theta}$ 
and $\Theta$ is the temperature of the system. $V_{Q\bar{Q}}$ is the potential between the heavy Quark and anti-Quark. The Polyakov loop is the trace of the Wilson line integrated from 0 to $\beta$ in Euclidean time, i.e., $P(x) = \frac{1}{N}Tr\Big [T exp(-i\int_0^{\beta}A_0(x,\tau)d\tau \Big ]$. Here, $A(x,\tau)$ is the pure gauge field, and $T$ denotes time ordering. The trace is over all the colors of the gauge field.
Based on AdS-CFT conjecture, the expectation value of the Polyakov loop correlator for a {\fontfamily{pzc}\selectfont N} $= 4$ supersymmetric Yang-Mills Lagrangian, could be equated to the string action = $S = \frac{1}{2\pi \alpha'}\int_{\Sigma}d\sigma d\tau \sqrt{G_{MN}\partial_{\alpha}X^M\partial_{\beta}X^N}$, 
%gans
where $\Sigma$ is the region between the Polyakov loops. While evaluating the Polyakov loop correlator, the Minkowski space time is converted to an Euclidean time by the use of imaginary periodic time.
%where $\Sigma$ is the domain covered by the Wilson loop, which lives at the Minkowski boundary of the $AdS_5$ space. 
$\Sigma$ corresponds to $\Sigma_1 + \Sigma_2$ in Fig. \ref{fig:single_loop}.
%In the integral, $\tau$ varies from $0$ to $\beta$, and $\sigma$ takes the value from $0$ to $L$. 

We calculate the above string action $S$ in the gravity dual domain, in Sec.~\ref{sec:tempgrad}.

In the imaginary time formalism, the spacetime fabric is seen as a topological (uniform) cylinder in a Euclideanized space of radius $\frac{\beta}{2\pi}$. The contour in Fig.~\ref{fig:single_loop} seems to indicate that the cylinder is non-uniform, with radii $\frac{\beta}{2\pi}$ and $\frac{\beta + \Delta \beta}{2\pi}$ at the two ends. However, an alternate viewpoint is that the topological cylinder now constitutes a uniform cylinder in a perturbed Euclidean space (similar to a geodesic being a straight line in curved space). Indeed, as we shall see going forward, in the dual AdS formulation, we perturb the AdS spacetime to model the temperature gradient. 
%gans: ref_rev1
%gans: ref_rev1
 If one were to evaluate the Polyakov loop correlator, in Eq.~\ref{eq:qqbeta}, in the field theoretic domain, one would then at least in principle, evaluate the correlator in the perturbed Euclidean space, with a "constant" $\beta$. In this work, we however evaluate it in the dual domain. We see that the variation in $\beta$ is now recast as a variation in the metric.

%%%%%%%%%%%%%%%%%%%%%%%%%%%%%%%%%%%%

\subsection{Finite Temperature in Dual Domain} 
%and the Temperature Gradient Calculation }
\label{sec:dualdomain}
%gans
Temperature is introduced in the AdS space, by introducing a black hole.
The black hole emits Hawking radiation, and creates a temperature, $\Theta$, in the system.
The resulting $AdS_5\times S_5$ near-horizon metric, of near extremal D3-branes in Type IIB string theory at finite temperature is given by~\cite{mald2, soo, witten}:
%end gans
%The $AdS_5\times S_5$ metric at finite temperature is given by~\cite{mald2, soo}
\begin{eqnarray}
\label{eq:base_metric}
\nonumber ds^2 = \alpha'\Big [ G \left ( -V dt^2 + dx_{||}^2 \right ) \\ 
+ \frac{1}{G}\left ( \frac{1}{V} du^2 + u^2 d\Omega_5^2 \right ) \Big ].
\end{eqnarray}
Here,
\begin{description}
\item $V = 1 - \frac{a^4}{u^4}$. %with $a^4 = \frac{2^7}{3}\pi^4g^2\mu$. The free energy density, $\mu$, is related to temperature $\Theta$ and is given by $\mu=\frac{4\pi^2}{45}\Theta^4$.
\item $G = \frac{u^2}{R^2}$,
\item $R^4 = 4\pi gN$, with $N$ large.
\end{description}
Imposing a periodicity condition on $\tau (= it)$ to prevent a conical singularity near the horizon, determines the temperature $\Theta$ \cite{witten}.

%gans
As discussed, when the thermal medium is having a temperature gradient, the Quark and anti-Quark can be at slightly different temperatures, $\Theta$ and $\Theta + \Delta \Theta$.
Thus the two particles exist at two different points in space, with slightly different metric, due to different temperatures. 
In the next section, Sec.~\ref{sec:black_string}, we show that it is possible to achieve a temperature gradient by inserting a non-uniform black string. The non-uniform black string metric can be obtained by perturbing the Schwarzchild black hole metric, namely,    
\begin{eqnarray}
\label{eq:mod_metric}
	\nonumber ds^2 = \alpha'\Big [  \left ( -GV +\Delta H_{tt}(t,u,x_{||}) \right ) dt^2 + Gdx_{||}^2  \\ 
\nonumber+  \big ( \frac{1}{GV}+\Delta H_{uu}(t,u,x_{||}) \big ) du^2\\
	+ \big ( \frac{u^2}{G} + \Delta H_{\Omega} \big ) d\Omega_5^2  \Big ].
% + \big ( u^2 + \Delta H_{\Omega} \big ) d\Omega_5^2 \Big ) \Big ].
\end{eqnarray}
where $G = \frac{u^2}{R^2}$.
The values of $\Delta H(t,u,x_{||})$ would be calculated in Sec.~\ref{sec:black_string}. It is shown how these perturbations lead to a temperature gradient. 
It is also shown that, after imposing periodicity conditions, under appropriate conditions, one can have $\beta$ varying linearly with $x_{||}$.
%\input{black_string}
%\section{Non Uniform Black string}
\section{Non Uniform Black string}
\label{sec:black_string}
\subsection{Perturbation of Black String} 
\label{sec:black_string_perturb}
In this section, we explore the incorporation of a temperature gradient in the AdS-CFT framework by inserting a non-uniform black string.
A non-uniform black string, in itself, need not necessarily result in a non-uniform temperature~\cite{bs}. 
We arrive upon a non-uniform black string, which gives rise to a temperature gradient, by perturbing a uniform black string. 
%gans:ref_rev1
%gans:ref_rev1
%gans:ref_rev2
We do not discuss the microscopic details of the source of perturbation, but only with the evolution of the perturbation. In the AdS space, the evolution would happen in vacuum, albeit with a negative cosmological constant. 
The perturbation can be caused by the black string capturing positive energy (or mass). Thus the energy momentum tensor corresponding to the positive energy can be expected to be physically viable and satisfy conditions like the average null energy condition.
%gans:ref_rev2
This section discusses the perturbation mechanism and the resultant temperature gradient. 

The $Q\bar{Q}$ pair is taken to lie in the spatial z-direction. This allows 
the 10 dimensional space to be first reduced to 8 dimensions by representing the spatial $x_{||}$ space defined by the three $x$, $y$ and $z$ dimensions, by the single $z$ dimension, for the purpose of analyzing black strings. 
A uniform black string in the AdS space can be modeled by using the AdS-Schwarzchild metric, as:
	%gans:ref_rev1
\begin{equation}
\label{eq:adssch}
%	\nonumber ds^2 = -V_Gdt^2 + \frac{du^2}{V_G} + \frac{u^2}{G}d\Omega_5^2 + Gdz^2,
	ds^2 = -V_Gdt^2 + \frac{du^2}{V_G} + \frac{u^2}{G}d\Omega_5^2 + Gdz^2,
\end{equation}
where $V_G = GV = G(1 - \frac{ a^4}{u^4})$, and G = $\frac{u^2}{R^2}$.

%gans: ref_rev1
%gans: ref_rev3
Before proceeding further, a note on terminology. 
The AdS-Schwarzchild solution is actually a black 3-brane, with $x_{||}$ in Eq. 4 being the 3 dimensional coordinate space parallel to the 3-brane. The Quark and anti-Quark are separated by a distance L in this space. We have called the axis aligned to $Q - \bar{Q}$ separation as z-axis.
The 1 dimensional z-axis subspace within the $x_{||}$ subspace is the only dimension really relevant to our analysis, and we have taken the temperature of the black 3-brane to also vary along the z-axis. Hence we are calling this as a black string which extends spatially along the z-axis, as the other two dimensions are irrelevant.
Additionally, along the u-axis, the Quark and anti-Quark lie at $u \rightarrow \infty$, as they are infinitely heavy.
%gans: ref_rev1

%gans: ref_rev1
We now apply a perturbation $\Delta H_{bc}$ to the metric in Eq.~\ref{eq:adssch}, which we now call $g_{bc}$, in order to model a non-uniform black string.
Motivated by the work done in ~\cite{greg}, we take the general form of $\Delta H_{bc}$ to be:\\
%e^{\Omega t}\sin(mz) 
%\begin{math}
\begin{eqnarray}
\nonumber \Delta H_{bc} = 
\\e^{\Omega t}e^{-mz} 
\left [
\begin{array}{cccccc}
h_{tt} & h_{tu} & 0 & 0 &... & h_{tz} \\
h_{ut} & h_{uu} & 0 & 0 &... & h_{uz} \\
0 & 0 & W & 0 &... & 0 \\
0 & 0 & 0 & W\sin^2(\phi_1) &... & 0 \\
. & . & . & . & ... & 0 \\
h_{zt} & h_{zu} & 0 & 0 &... & h_{zz} \\
\end{array}
\right ].
%\end{math}
\end{eqnarray}
%\end{table}
%gans: ref_rev1
\\
Here, $h_{tt}, h_{uu}, h_{ut}, h_{tu}$ and $W$ are some functions of $u$. The perturbation $\Delta H_{bc}$ satisfies the equation $\Delta_L(\Delta H_{bc}) = 0$, where $\Delta_L$ is the Lichnerowicz operator. 
%gans: ref_rev1
This is a linearized approximation of gravity, applicable for small perturbations of the metric.
%gans: ref_rev1
%gans: ref_rev2
At a simplistic level, a small perturbation may be defined as $|\Delta H_{bc}| \ll |g_{bc}|$. While, this definition works for diagonal elements $h_{tt}$, $h_{uu}$, $W$ and $h_{zz}$, this definition does not work for non diagonal elements like $h_{tu}$ and $h_{ut}$ as the corresponding $g_{tu}$ and $g_{ut}$ are 0. 
However, as we shall later see in Eq.~\ref{eq:ABC}, $h_{tu}$ and $h_{ut}$ are not independent of $h_{tt}$ and $h_{uu}$. Hence, once $h_{uu}$ and $h_{tt}$ are taken small in the sense of $|\Delta H_{bc}| \ll |g_{bc}|$, $h_{tu}$ and $h_{ut}$ automatically get fixed.
A detailed discussion on the magnitude of perturbation is given in Sec.~\ref{sec:magpert}.
%Additionally, if the metric perturbation is small, it would mean that the change in the black string horizon radius and consequently, the change in black string mass per unit length is small.
%gans: ref_rev2

Taking a transverse gauge, $\nabla_b h^b_c - \frac{1}{2}\nabla_ch = 0$, the equation, $\Delta_L(\Delta H_{bc}) = 0$, reduces to~\cite{greg}:
\begin{equation}
\label{eq:per}
\nabla^2 (\Delta H_{bc}) + 2R_{bdc}^e(\Delta H^d_e) = 0,
\end{equation}
with $R_{bdc}^e$, being the Riemann curvature tensor.
%%%%%%%%%new %%%%%%%%
In ~\cite{greg}, the $z$ dependence has been taken as $e^{imz}$. 
It  can be seen that, with $e^{imz}$, the value of $m$ which would solve the Lichnerowicz equation for our chosen value of $\Omega$, is imaginary. 
Hence, from hindsight, we take the $z$ dependence as $e^{-mz}$ with real and positive $m$. A real physical black string is more likely to be of the form $e^{-m|z|}$. But such a function is not well behaved at $z=0$. Since we are mainly interested in the region, $z>0$, we have preferred to proceed with the more well behaved $e^{-mz}$  instead of $e^{-m|z|}$. The two functions have the same value in the region of interest $0<z<L$.
%%%%%%%%%%%%%%%%%

As we are interested in a particular solution, and not in the most general solution, we take, $h_{zz}=h_{tz} = h_{zt} = 0$.  
From Eq.~\ref{eq:per}, one can then get  the following set of differential equations: 
\begin{eqnarray}
\label{eq:dalem1}
\nonumber  \Big [ V_Gh''_{tt} + (\frac{3V_G G'}{2G} - V_G')h'_{tt} - V_G''h_{tt}\\ 
\nonumber+ \left ( \frac{h_{tt}}{2V_G} - \frac{V_G}{2}h_{uu} \right ) V_G'^2 \\
\nonumber + \left (- \frac{3 h_{tt}G'}{2G} + \Omega h_{ut} + \Omega h_{tu}  \right )V_G'\\
%changing the sign of m^2 due to e^{-mz} instead of sin(mz)
%- \left ( m^2 + \frac{\Omega^2}{V} \right ) h_{tt}  \\ 
\nonumber - \left ( -\frac{m^2}{G} + \frac{\Omega^2}{V_G} \right ) h_{tt}  \\ 
-  \left ( V_G^2 h_{uu}V_G'' \right ) \Big ] = 0,
\end{eqnarray}

\begin{eqnarray}
\label{eq:dalem2}
\nonumber \Big [ V_G h''_{uu} + \left (3V_G' + \frac{3V_G G'}{2G}\right ) h'_{uu} + h_{uu}V_G'' \\
\nonumber + \left ( \frac{h_{uu}}{2V_G} - \frac{h_{tt}}{2V_G^3} \right ) V_G'^2\\
\nonumber + \left ( \frac{3 h_{uu}G'}{2G} + \frac{\Omega}{V_G^2} (h_{tu} + h_{ut}) \right ) V_G'\\
%changing the sign of m^2 due to e^{-mz} instead of sin(mz)
%\nonumber + \left ( -\frac{-10V}{u^2} - \left( m^2 + \Omega^2/V \right )  \right ) h_{uu}\\ 
	\nonumber + \left (  \frac{m^2}{G} - \frac{\Omega^2}{V_G}  \right ) h_{uu}\\ 
-  \frac{V_G'' h_{tt}}{V_G^2}  \Big] = 0,
\end{eqnarray}
\begin{eqnarray}
\label{eq:dalem3}
\nonumber \Big [ V_G h_{tu}'' + (V_G' + 3V_G\frac{G'}{2G})h_{tu}'\\
\nonumber - (1/2)(\frac{h_{tu}}{V_G} + \frac{h_{ut}}{V_G})V_G'^2\\
\nonumber + \Omega(h_{uu} + \frac{h_{tt}}{V_G^2})V_G' - \frac{3}{4}h_{tu}(\frac{G'}{G})^2 \\
+ (\frac{m^2}{G} - \frac{\Omega^2}{V_G})h_{tu} - h_{tu}V_G'' \Big ] = 0,
\end{eqnarray}

\begin{eqnarray}
\label{eq:dalem4}
\nonumber \Big [  W'V_G' + V_GW'' + \frac{3V_G G'}{2G}W'\\
	\nonumber +   \left ( \frac{m^2}{G}W - \frac{\Omega^2}{V_G}W \right )\\
  - \frac{8W}{R^2} \Big ] = 0.
\end{eqnarray}

The gauge condition, $\nabla_bh^b_c - \frac{1}{2}\nabla_ch=0$, leads to 
\begin{eqnarray}
\label{eq:gauge1}
\nonumber V_Gh'_{ut} + V_G'h_{ut} + \frac{3V_G G'}{2G}h_{ut} - \frac{\Omega}{V_G}h_{tt} \\
- \frac{1}{2}h_{ut}V_G' + \frac{1}{2}h_{tu}V_G' - \frac{\Omega}{2}h= 0,
\end{eqnarray}
and
\begin{eqnarray}
\label{eq:gauge2}
\nonumber V_Gh'_{uu} + V_G'h_{uu} + \frac{3V_G G'}{2G}h_{uu} - \frac{\Omega}{V_G}h_{tu} \\
+ \frac{1}{2}h_{uu}V_G' + \frac{h_{tt}V_G'}{2V_G^2} - \frac{1}{2}h' = 0.
\end{eqnarray}
%%%   The final gauge condition, $h^b_b = 0$, gives 
%%%   \begin{equation}
%%%   \label{eq:gauge3}
%%%   \left ( \frac{-h_{tt}}{V_G} + h_{uu}V_G \right ) + 5W = 0.
%%%   \end{equation}

The superscript $'$ refers to partial derivative w.r.t $u$. 
The region of interest is the near horizon region, as $u\rightarrow a$. Let $u_+ = u - a$, and in the region of interest, $u_+ \ll u$. In these limits, the lowest power of $u_+$ becomes the most significant. 
We further assume, that, the solution in the lowest power of $u_+$, to be of the form, $h_{tt}(u) = A(m,\Omega,a)u_+^n$, $h_{uu}(u) = B(m,\Omega,a)u_+^l$, $h_{ut}(u) = h_{tu}(u) = C(m,\Omega,a)u_+^q$ and $W = W_0(m,\Omega,a) u_+^k$.

For the above form of perturbation functions, a nontrivial solution is obtained if, $n = l+2$, $q=n-1$ and $k = n-1$. 
%A lower power of $q$ than $n-1$, gives C=0, while solving the above equations.
Since the lowest power of $u_+$ becomes dominant as $u_+ \rightarrow 0$, by equating the coefficients of the lowest power of $u_+$ to 0, the equations \ref{eq:dalem1},~\ref{eq:dalem2},~\ref{eq:dalem3},~\ref{eq:dalem4},~\ref{eq:gauge1} and \ref{eq:gauge2} give respectively,
%gans: ref_rev1
\begin{eqnarray}
	\nonumber \frac{4Aa}{R^2}(n^2 - 2n + 0.5) - \frac{1}{2}(\frac{4a}{R^2})^3B\\
\nonumber + 2\Omega C\frac{4a}{R^2} - \Omega^2 A\frac{R^2}{4a} = 0,\\
\\
\nonumber B\frac{4a}{R^2} ( l^2 + 2l + 0.5) - A\frac{1}{2}\frac{R^2}{4a} + 2\Omega C \frac{R^2}{4a}\\
\nonumber - B\Omega^2\frac{ R^2 }{4a} = 0,\\
\\
\nonumber \frac{4a}{R^2}C(q^2 - 1) + \Omega\left ( B + A\left (\frac{R^2}{4a}\right )^2 \right )\frac{4a}{R^2}\\
\nonumber - \Omega^2C \frac{R^2}{4a} = 0,\\
%- \Omega^2C \frac{R^2}{4a} = 0,\\
\\
%\end{eqnarray}
%\begin{eqnarray}
%\nonumber \\
%equation for W
\nonumber k^2W_0\frac{4a}{R^2} - \Omega^2W_0\frac{R^2}{4a} =0,\\
\\
%\end{eqnarray}
%\begin{eqnarray}
%\nonumber C\frac{4a}{R^2}(q+1) - A\frac{\Omega R^2 }{4a} = 0,\\
\nonumber C\frac{4a}{R^2}(q+1) - A\frac{\Omega R^2 }{4a}\\ 
%gauge condition: added -1/2\Omega h
\nonumber -\frac{\Omega}{2} \left (-\frac{AR^2}{4a} + \frac{4Ba}{R^2} +\frac{5W_0}{R^2} \right ) = 0,\\
\\
\nonumber B\frac{4a}{R^2}(l + 3/2) - C\frac{\Omega R^2 }{4a} + A\frac{1}{2}\frac{R^2}{4a}\\ 
%new gauge condition: added -1/2h'
\nonumber -\frac{1}{2}\left ( -(n-1)\frac{AR^2}{4a} + (l+1)\frac{4Ba}{R^2} + k\frac{5W_0}{R^2} \right )= 0.\\
\end{eqnarray}
%gans: ref_rev1

The above equations are satisfied for $n=1$, $l = -1$, $q = 0$ and $k=0$, with the constraint that $W_0=0$ and $\Omega\frac{ R^2}{4a} = \pm 1$. 
In ~\cite{greg}, the authors have solved the equations for arbitrary values of $\Omega$, but in this work, we have restricted ourselves to certain values of $\Omega$, since it gives rise to a form of metric perturbation, which is more suitable for temperature gradient calculations.
If $R=a$, then, it is seen that our solution corroborates with ~\cite{greg} for the specific values of $\Omega = \frac{4}{a}$, when $h_{tu} \propto u_+^{-1 + \frac{\Omega a}{4}}$ and $\Omega = -\frac{4}{a}$ when $h_{tu} \propto u_+^{-1 - \frac{\Omega a}{4}}$.
Further, we get 
\begin{eqnarray}
\label{eq:ABC}
\nonumber C = A\Omega(\frac{ R^2 }{4a})^2,\\ 
	B = A(\frac{R^2 }{4a})^2.~~
\end{eqnarray}

%Inserting the value of $B$ in the gauge condition, $h^b_b=0$ (Eq.~\ref{eq:gauge3}), we find that $W$ vanishes. 
The non zero contribution to $W$ comes from equating with higher powers of $u_+$.
%%%   To determine $W$, we then solve Eq. ~\ref{eq:dalem3}, to obtain,
%%%   \begin{equation}
%%%   \left ( \frac{4k^2W_0}{a} - \frac{\Omega^2 a W_0}{4} \right ) u_+^{k-1} + higher~powers~of~u_+ = 0.
%%%   \end{equation}
%%%   In the limit $u_+ \rightarrow 0$, we get,
%%%   \begin{math} 
%%%   k = \pm \frac{\Omega a}{4}.
%%%   \end{math} 
%%%   Thus, 
%%%   \begin{equation}
%%%   \label{eq:Weq}
%%%   5W = -(-\frac{h_{tt}}{V} + Vh_{uu}) \approx 5W_0 u_+^{\pm\frac{\Omega a}{4}},
%%%   \end{equation} 
%%%   which then corroborates with the solution of $-\frac{h_{tt}}{V} + Vh_{uu}$ in ~\cite{greg_book}.

However, for the purpose of determining the temperature of the perturbed black string, near the horizon, we are mainly interested in the dominant, lowest power of $u_+$ behavior of $h_{tt}$ and $h_{uu}$. 
Thus, with $n=1$, and $l=-1$, we get, $h_{tt} = Au_+$, and $h_{uu} = \frac{A}{u_+}\left (\frac{R^2}{4a}\right )^2$, along with the value of $\Omega = \pm \frac{4a}{R^2}$.
For large $R$, $\Omega$ is small, indicating a slow time varying system, as required for a system in local thermodynamic equilibrium.
For clarity, we summarize the final form of $\Delta H_{tt}$, $\Delta H_{u_+u_+}$ and $\Delta H_{tu_+}$ as:
\begin{eqnarray}
\nonumber \Delta H_{tt} = e^{\Omega t}e^{-mz}Au_+,\\
\nonumber \Delta H_{u_+u_+} = e^{\Omega t}e^{-mz}B/u_+,\\
\nonumber \Delta H_{tu_+} = e^{\Omega t}e^{-mz}C.
\end{eqnarray}

The perturbed black string metric now becomes: 
\begin{eqnarray}
%gans
\label{eq:temp_metric}
\nonumber ds^2 = (-V_G + e^{\Omega t}e^{-mz}Au_+)dt^2 \\
\nonumber + \left ( \frac{1}{V_G}+\frac{e^{\Omega t}e^{-mz}B}{u_+} \right ) du_+^2\\
\nonumber + \left ( e^{\Omega t}e^{-mz}C \right ) dtdu_+\\
 + (...)d\Omega_5^2 + Gdz^2,
\end{eqnarray}
where we have replaced $du$ by $du_+$.

In order to eliminate the $dtdu_+$ term, we apply a linear transformation $t' \leftarrow t + \alpha u_+$, and $u_+' \leftarrow u_+$. We obtain the metric,
\begin{eqnarray}
\nonumber ds^2 = \left (-V_G + e^{\Omega t'}e^{-mz}A u'_+\right ) dt'^2 \\
\nonumber + \Big ( \frac{1}{V_G}+\frac{e^{\Omega t'}e^{-mz}B}{u'_+} \\
\nonumber - \frac{e^{2\Omega t'}e^{-2mz} C^2a}{4(-V_G + e^{\Omega t'}Aae^{-mz}u_+)} \Big ) du'^2_{+}\\
%- \frac{e^{2\Omega t}\sin^2 (mz) C^2a}{4(-4 + e^{\Omega t'}Aa\sin(mz)} \Big ) du'_+^2\\
+ (...)d\Omega_5^2 + Gdz^2.
\end{eqnarray}
In the above equation, we have used the fact that, as $u_+ \rightarrow 0$, $u_+' \rightarrow 0$ and $e^{\Omega (t' - \alpha u_+)} \rightarrow e^{\Omega t'}$.  
We take the inter Quark distance $L \ll 1/m$ along the z-axis. In Sec. ~\ref{sec:num_sims}, we shall see that, this would be true if $R$ is very large, leading to a small $m$. In this region of the black string, i.e., $0 < z < L$, we can approximate $e^{-mz}~\approx~(1-mz)$. 
In the dual domain, these assumptions become equivalent to the slope of the temperature gradient being very small.
    With these approximations, and with $C^2 < B$, and dropping the primes from $t'$ and $u'_+$, we get the first two terms of the metric as,
%gans: ref_rev1
\begin{eqnarray}
\label{eq:twoterms}
	\nonumber ds^2_{2 terms} = \left (-V_G + e^{\Omega t}(1-mz)A u_+ \right ) dt^2 \\
+ \left ( \frac{1}{V_G}+\frac{e^{\Omega t}B(1-mz)}{u_+} \right ) du^2_+.
\end{eqnarray}
The term $ds^2_{2 terms}$ refers to only the first two terms of the metric $ds^2$ being considered.
%gans: ref_rev1

 It would be tempting to relate the evolution, $e^{\Omega t}$, with the evolution of the temperature of the thermal medium.
If $\Omega = -\frac{4a}{R^2}$, it indicates that after infinite time, the perturbation to the metric vanishes. This qualitatively agrees with the fact that the temperature gradient in a time evolving thermal medium like the QGP system should also vanish after sufficiently long time.
Its also possible that a different function $f(t)$ (say, for example $f(t)$ = $\sum_i c_ie^{\Omega_i t}$) may more correctly determine the evolution. We, however, in this work, do not probe this further.
Relating the evolution $e^{\Omega t}$, with the evolution of a thermal medium like the QGP, however has a drawback that the evolution lifetime of QGP is finite, while the metric evolution $e^{\Omega t}$ decays over infinite time. We premise that the metric is the dual to the thermal medium only till the QGP lifetime, $t_{QGP}$. 

We now analyze the system over a small time interval $\Delta t = t_2 - t_1$.
%We split the $e^{\Omega t}$ into a time averaged $F_1$ (averaged over $t_{QGP}$), and a time dependent $F_2(t)$ part. 
We split the $e^{\Omega t}$ into a time independent $F_1$ (averaged over $\Delta t$), and a time dependent $F_2(t)$ part. 
%Actually, as discussed in Sec.~\ref{sec:intro}, the time interval of averaging can be any interval greater than the time of the mean free path of the particles constituting the thermal system.
The first two terms of the metric in Eq.~\ref{eq:twoterms} then become 
%gans: ref_rev1
\begin{eqnarray}
	\nonumber ds^2_{2 terms} = \left (-V_G + (F_1 + F_2(t))(1-mz)A u_+ \right ) dt^2 \\
\nonumber + \left ( \frac{1}{V_G}+\frac{(F_1 + F_2(t))B(1-mz)}{u_+} \right ) du^2_+,\\
\end{eqnarray}
%gans: ref_rev1
where 
\begin{itemize}
%\item $F_1 = (e^{\Omega t_{QGP}} - e^{\Omega t_1})/(\Omega(t_{QGP} - t_1))$.
\item $F_1 = (e^{\Omega t_{2}} - e^{\Omega t_1})/(\Omega(t_{2} - t_1))$.
\end{itemize}
and $t_1$ and $t_2$ corresponding to the beginning and end of the time interval $\Delta t$.

As mentioned in Sec.~\ref{sec:intro}, we are attempting to model $\Theta_0(z)$ in Eq.~\ref{eq:tempgrad_def}.
%gans
We use a quasi-stationary approach, the justification for which is provided in appendix ~\ref{sec:quasi} for both the primal and dual domain. The quasi-stationary approach becomes more appropriate, when the time interval of integration, $\Delta t$, to determine $F_1$, is small.
The full time varying metric would be dual to the time varying thermal system. 
For a slow time varying system, and for small time interval $\Delta t$ (i.e, when the condition in Eq.~\ref{eq:approxcond} is satisfied), $\Delta \Theta(z,t)$ in Eq.~\ref{eq:tempgrad_def} is small. Similarly $F_2(t)$ part of the metric would be small if the rate of time evolution of the metric is small. This means that the error in considering only the static part of the system in either the primal or dual domain would be small. 
Thus, under the above circumstances, neglecting the small time dependent part, we use the time independent part of the metric to represent or roughly be the dual of the time independent thermal system, albeit with some error. 
%and the time dependent part of the metric to model the duality of the time dependent QGP system. 
%gans
The first two terms of the metric corresponding to the time independent QGP system would then be

%gans: ref_rev1
\begin{eqnarray}
	\nonumber ds^2_{2 terms} = \left (-V_G + F_1(1-mz)A u_+ \right ) dt^2 \\
+ \left ( \frac{1}{V_G}+\frac{F_1 B(1-mz)}{u_+} \right ) du^2_+.
\end{eqnarray}
%gans: ref_rev1

Near horizon, as $u\rightarrow a$, $V_G = \frac{u^2}{R^2}(1 - \frac{a^4}{u^4}) \approx 4au_+/R^2$. 
Setting $\rho = \sqrt{u_+}$, the first two terms become:
%gans: ref_rev1
\begin{eqnarray}
\label{eq:metric}
%\nonumber \left ( \frac{1}{4a}( R^2 + 4aF_1Be^{\Omega t}(1-mz) \right ) \\
%new_journal
	\nonumber ds^2_{2 terms} = \frac{1}{4a}\left (  R^2 + 4aF_1B(1-mz) \right ) \\
\nonumber	\times \left [ \rho^2 \left ( \frac{4a}{R^2} \right )\frac { \left ( -4a + R^2 F_1 (1-mz)A \right ) }{\left( R^2 + 4aBF_1(1-mz) \right )} dt^2 +  d\rho^2 \right ].\\
\end{eqnarray}
%gans: ref_rev1

We are now at a point where we can define the temperature. 
With $\beta = it$, and taking the periodicity condition to avoid conical singularity, we get,
\begin{math}
\sqrt{\left ( \frac{4a}{R^2} \right ) \frac {\left ( -4a + R^2 F_1 (1-mz)A \right ) }{\left( R^2 + 4aBF_1(1-mz) \right )}}\beta = 2\pi,
\end{math}
giving
\begin{equation}
\label{eq:nonlintemp}
\beta = 1/\Theta = 2\pi \sqrt{\left ( \frac{R^2}{4a} \right ) \frac {\left( R^2  + 4aBF_1(1-mz) \right )}{ \left ( -4a + R^2 F_1 (1-mz)A \right )}}.
\end{equation}
Thus we have obtained $\beta=1/\Theta$ as a function of $z$.

If, we consider very small perturbations, i.e., B and A are small, $\beta$ can be seen to be approximately a linear function of $z$.
%As discussed earlier, in the region of interest, value of $mz$ is also small. 
If $\frac{4a}{R^2} \gg F_1 e^{-mz}A \approx F_1 \,(1-mz)\,A$, then we get the full metric as: 
%gans: ref_rev1
\begin{eqnarray}
%\nonumber ds^2 \approx \left (\frac{1}{4a}( R^2 + 4aF_1Be^{\Omega t}(1-mz) \right ) \\
%new journal
\nonumber ds^2 \approx \frac{1}{4a}\left ( R^2 + 4aF_1B(1-mz) \right ) \\
	\nonumber \times \Big [ (\frac{4a}{R^2})^2 \frac {-\rho^2}{\left \{1 + \left (\frac{4a}{R^2}F_1 B + AF_1 (\frac{R^2}{4a})\right )(1-mz) \right \} } dt^2\\
\nonumber + d\rho^2 \Big ] + (...)d\Omega_5^2 + dz^2.\\
\end{eqnarray}
%gans: ref_rev1
We rewrite the metric as:
\begin{eqnarray}
%new journal
%\nonumber ds^2 = \left (\frac{1}{4a}( R^2 + 4aF_1Be^{\Omega t}(1-mz) \right ) \\
\nonumber ds^2 = \frac{1}{4a}\left ( R^2 + 4aF_1B(1-mz) \right ) \\
\times \left [ \frac {(4a/R^2)^2 }{(1 + Emz)} \rho^2 d\beta^2 + d\rho^2 \right ]
 + (...)d\Omega_5^2 + dz^2,
\end{eqnarray}
with $E~=~-(\frac{4a}{R^2}F_1 B + AF_1 (\frac{R^2}{4a}))~=~ -2AF_1 \frac{R^2}{4a}$, and since the perturbations are small, $|E| \ll 1$. 
%Additionally, we have taken $\beta = it$.
Taking the periodicity, 
\begin{eqnarray}
\label{eq:periodicity}
\frac{4a\beta}{R^2\sqrt{1 + E\,mz}} = 2\pi,
\end{eqnarray}
we get,
\begin{eqnarray}
\label{eq:linear_tempgrad}
\beta = \frac{1}{\Theta} = \frac{\pi R^2}{2a}\sqrt{1 + E\,mz} \approx \frac{\pi R^2}{2a}(1 + \frac{Em}{2}z).
\end{eqnarray}
This gives temperature = $\Theta$ = $1/\beta$ as a function of $z$. Since the Quark and anti-Quark lie along the $z$ axis, The temperature can be seen varying along the Quark anti-Quark distance, leading to a temperature gradient. The inverse temperature gradient = $\frac{d\beta}{dz}$ = $\pi R^2Em/4a$. The linear gradient is a valid assumption only when the perturbation $E$ and $m$, are small. For larger perturbation, there would still be a temperature gradient along the Quark anti-Quark axis, but the gradient would not be linear.

It can be seen that a large $R$ corresponds to a large $\beta$ and a small $\Omega$, justifying both the assumptions of a large $\beta$ and a pseudo static system (for a system in local thermodynamic equilibrium). In Sec.~\ref{sec:num_sims}, we will see that $m= 14a/R^2$, leads to a solution of Lichnerowicz equation for $\Omega=-4a/R^2$. Hence, a large value of $R$ also leads to a small value of $m$, and thus consistent with the assumptions used in our derivation. 
 
%\section{Numerical sims}
%\appendix
\subsection{Numerical Simulation}
\label{sec:num_sims}
We now fully solve the Lichnerowicz equation and show that the solution is bounded for the value of $\Omega=-4a/R^2$.
%%%   The equation for $h_{tu}$, with all the other variables eliminated, has been provided in \cite{greg}. As $u_+\rightarrow 0$, taking the lowest power of $u_+$, the equation reduces to 
%%%   \begin{eqnarray}
%%%   \nonumber \left ( -\Omega^2 + \frac{(D-3)^2}{4a^2}\right ) h_{tu}''\\
%%%   \nonumber - 3(D-3)\left (\frac{\Omega^2}{aV} + \frac{(D-3)^2}{4a^3V}\right ) h_{tu}' +\Big [ \left(\frac{\Omega^2}{V}\right)^2 \\
%%%   - (D-3)^2\left ( \frac{-5\Omega^2}{4a^2V^2} + \frac{(D-3)^2}{4a^4V^2} \right ) \Big ] h_{tu} = 0,
%%%   \end{eqnarray}
%%%   where, $D+1$ = $8$ = black string dimensions. 
%%%   It is easily seen that $h_{tu} = Cu_+^q$, with $q = 0$, solves the above equation for D=7 and $\Omega = \pm 4/a$.

We now numerically solve for $h_{tu}$, $h_{uu}$ and $h_{tt}$, for all $u$, using Eqs.~\ref{eq:dalem1}, ~\ref{eq:dalem2} and ~\ref{eq:dalem3}, along with the gauge conditions, ~\ref{eq:gauge1} and ~\ref{eq:gauge2}. They have been solved for $\Omega  = -4a/R^2$ and $m = 14a/R^2$, using matlab's ode15s function. The values of both "RelTol" and "AbsTol" are $1e^{-6}$.
%For the initial conditions, i.e. values as $u_+ \rightarrow 0$, we have taken the values accurate till the next higher order in $u_+$. Explicitly, 
For the initial conditions, i.e. values as $u_+ \rightarrow 0$, we have taken the values of A, B and C as discussed earlier.
%%%   \begin{eqnarray}
%%%   \nonumber h_{tt}(u_+\rightarrow 0)=Au_+ + A_2 u_+^2,\\
%%%   \nonumber h_{uu}(u_+\rightarrow 0)=B/u_+ + B_2,\\
%%%   h_{tu}(u_+\rightarrow 0)=C + C_2u_+,
%%%   \end{eqnarray}
%%%   where 
%%%   $C_2 = \frac{-3m^2 a}{20}A, B_2= \frac{-m^2}{\Omega 40}a^2A$ and $A_2 = \frac{10}{a}A$. 

From Eq.~\ref{eq:ABC}, $A = \frac{C}{\Omega}(\frac{4a}{R^2})^2$, and we take $C$ = 0.1 as reference. The differential equations have been solved from $u_+$ = $1e^{-5}$ to $9a$ ($\equiv$ to $u$ varying from $(a+1e^{-5})$ to $10a$), along a logarithmically spaced grid with $10^6$ points.

\begin{figure}[h!]
\includegraphics[width = 80mm,height = 80mm]{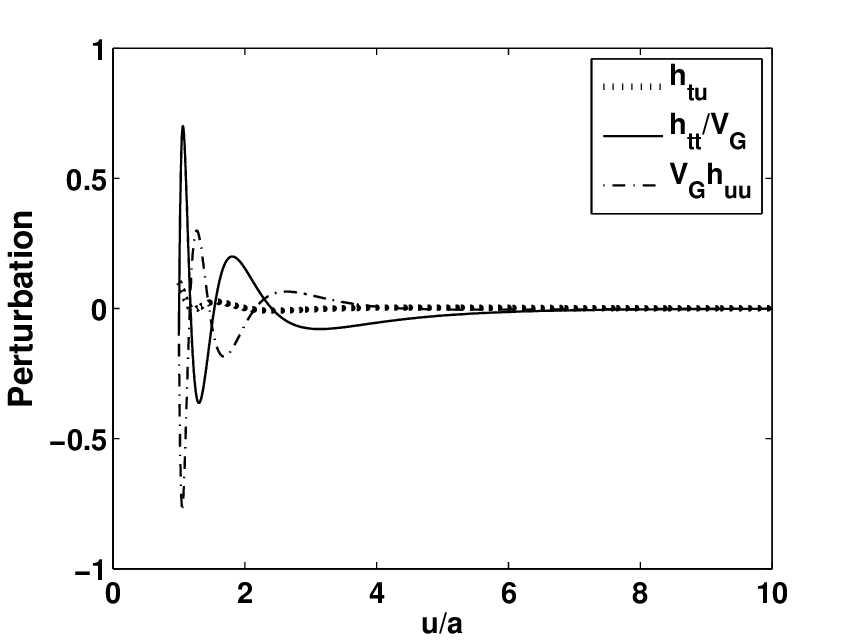}
\captionsetup{justification=raggedright, singlelinecheck=false}
\caption{Solution for $h_{tu}$, $\frac{h_{tt}}{V_G}$ and $V_Gh_{uu}$ with $\Omega$ = -$4a/R^2$, and $m$ = $14a/R^2$.}
\label{fig:hall}
\end{figure}

Figure~\ref{fig:hall} depicts the solution, with $\Omega$ = $-\frac{4a}{R^2}$ and $m$ = $\frac{14a}{R^2}$. The value of $a=R=20$.
It is seen that $h_{tt}(u)/V_G$, $V_Gh_{tu}(u)$ and $h_{tu}(u)$ are bounded for all values of $u$.
Evidently, $e^{\Omega t}e^{-mz}h_{tt}/V_G$, $e^{\Omega t}e^{-mz}V_Gh_{uu}$ and $e^{\Omega t}e^{-mz}h_{tu}$ would be bounded for all values of $t$ and $u$, and for $z>0$. 
Hence, we have arrived at a solution of the Lichnerowicz equation, with value of $\Omega =-4a/R^2$, and $m = 14a/R^2$, which is never going to blow up for positive $z$.

\begin{figure}[h!]
\includegraphics[width = 80mm,height = 80mm]{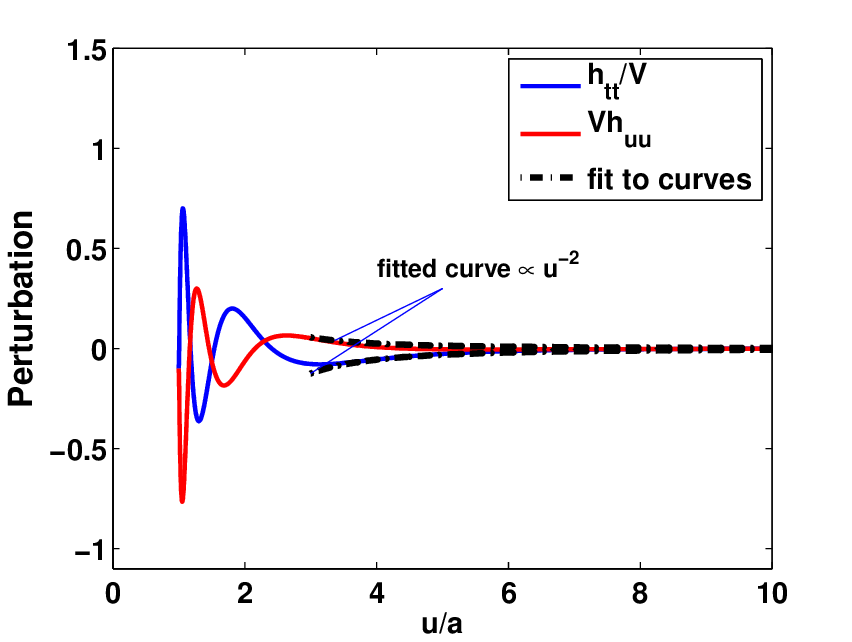}
\captionsetup{justification=raggedright, singlelinecheck=false}
\caption{Fitting a curve for large $u$, for $\frac{h_{tt}}{V}$ and $Vh_{uu}$ with $\Omega$ = -$4a/R^2$, and $m$ = $14a/R^2$.}
\label{fig:curvefit}
\end{figure}

To get an estimate of the large $u$ behavior of the perturbation, we fit a curve as shown in Fig.~\ref{fig:curvefit}. At large $u$, the curves seem to approximately follow $\frac{1}{u^{2}}$. 
%In Fig.~\ref{fig:curvefit}, the simulation is performed till $\frac{u}{a} = 20$.

\subsection{Magnitude of Perturbation}
\label{sec:magpert}
This section explores further the magnitude of perturbation for all $u$ and $t$. Large perturbations can lead to divergence of the integrals in the calculation of the action.
%new:gans
Near horizon, $g_{tt} + \Delta H_{tt} = V_G + e^{\Omega t_0}e^{-mz}Au_+ \approx (\frac{4a}{R^2} + e^{\Omega t_0} e^{-mz}A) u_+$. 
In Sec.~\ref{sec:black_string_perturb}, for linearity to be valid, we had assumed that A is small enough, such that $A \ll 4a/R^2$. 
Thus, the perturbation is small near the horizon for $g_{tt}$. 
Again, near horizon, $g_{uu} + \Delta H_{uu} = \frac{1}{V_G} + \frac{e^{\Omega t_0}e^{-mz}B}{u_+} \approx (\frac{R^2}{4a} + e^{\Omega t_0} e^{-mz}B) \frac{1}{u_+}$. As $u_+ \rightarrow 0$, the perturbation and the original metric, $g_{uu}$, approach infinity. Here, we need to define "a small perturbation". When we say that the perturbation is small, we do not mean $\Delta H_{uu}$ itself, but the coefficient of $1/u_+$ in $\Delta H_{uu}$, compared to the coefficient of $1/u_+$ in $g_{uu}$. 
From Eq. \ref{eq:ABC}, $B=A(\frac{R^2}{4a})^2$. Thus, $g_{uu} + \Delta H_{uu} = (\frac{R^2}{4a})^2\left ( \frac{4a}{R^2} + e^{\Omega t_0}e^{-mz}A \right )\frac{1}{u_+}$.

We now show that the perturbation remains small for all $u$ and any $t=t_0$.
From Sec.~\ref{sec:num_sims} and Fig.~\ref{fig:hall}, it can be seen that the solution for the black string perturbation, namely, $h_{tt}/V$ and $Vh_{uu}$ are bounded for all values of $u$, and in fact, roughly decaying with $u$. 
We identify the $z$ variable in this section with $x_{||}$ in Eq.~\ref{eq:mod_metric}, and understand that it lies in the range $0<x_{||}<L$.
The negative values of $\Omega$, namely, $\Omega = \frac{-4a}{R^2}$, indicates that if the perturbation is small at $t=0$, it would be small at any $t=t_0$.
Summarizing, the perturbation to the metric remains smaller than the original metric in the region of interest $0<x_{||}<L$, for all $u$, and for all $t_0$.

For a positive value of $\Omega = 4a/R^2$, the existence of Lichnerowicz solution leads to an unstable black string, as the solution blows up~\cite{greg}. In a real astrophysical black string, both positive and negative values of $\Omega$ would be possible, leading to the instability. 
However, in the QCD domain, the temperature or the temperature gradient of the QGP is not going to blow up, but rather decay with time. Hence we identify the negative value of $\Omega$, namely, $\Omega=-4a/R^2$, to model the appropriate mathematical dual.
%\section{Temperature Gradient Calculation}
\section{$Q\bar{Q}$ Potential under a Temperature Gradient}
\label{sec:tempgrad}
\subsection{Formulation}
\label{sec:tempgrad_formulation}
\begin{figure}[h!]
\includegraphics[width = 80mm,height = 80mm]{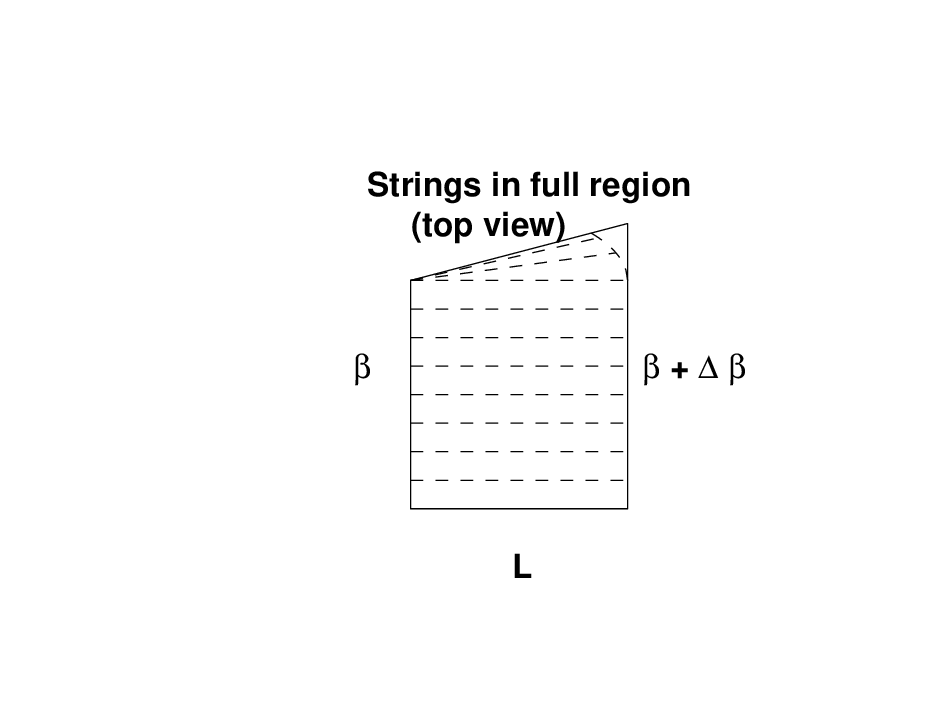}
\captionsetup{justification=raggedright, singlelinecheck=false}
%gans
\caption{ Dashed lines represent the string contours (top view) in the entire region between the Polyakov loops, as the string traverses the worldsheet area. $\beta$ and $\beta + \Delta \beta$ correspond to $\frac{1}{\Theta + \Delta \Theta}$ and $\frac{1}{\Theta}$ respectively.}
\label{fig:strings_full}
\end{figure}

We have so far derived the perturbation functions in the metric in Eq.~\ref{eq:mod_metric}, which can be used to calculate the Quark anti-Quark potential, in the presence of a temperature gradient.  We have seen that $\beta$ can vary linearly with $z$ (or equivalently $x_{||}$).
%gans
%The Wilson loop, corresponding to a linear variation of $\beta$ is shown in Fig.~\ref{fig:strings_full}, which is what we require to compute.
The string contours corresponding to the Polyakov loop correlator, for varying $\beta$ is shown in Fig.~\ref{fig:strings_full}, which is what we require to compute.
One can identify $\beta$ and $\Delta \beta$ in Fig.~\ref{fig:strings_full}, with black string metric parameters in Eq.~\ref{eq:linear_tempgrad}, at $z=L$, as:
\begin{eqnarray}
\label{eq:beta}
\nonumber \Delta \beta = \frac{\pi R^2 EmL}{4a},\\
\beta = \pi R^2/2a.~~~~~~~
\end{eqnarray}

The problem has now been reduced to mathematical computation of the string action corresponding to the string contours in Fig.~\ref{fig:strings_full}, using the metric in Eq.~\ref{eq:mod_metric}.
The general approach would require a numerical approach. However, many parts of the solution can be analyzed analytically.
As mentioned in \cite{soo}, beyond a certain inter Quark distance $d_{max}$, the free energy of strings are minimized when the strings are disconnected and run from the boundary to the black hole horizon. This configuration would correspond to a dissociated heavy Quarkonium. Since we are mainly interested in a bounded $Q\bar{Q}$ potential, we only consider the case where $L<d_{max}$. 

For  convenience, we redefine $h^r_{tt} \leftarrow \frac{R^2}{u^2}h_{tt}$, and similarly, $h^r_{uu} \leftarrow \frac{u^2}{R^2}h_{uu}$, where the superscript $r$ stands for "redefined". With this redefinition, the static part of the perturbed metric can be written as: 
%gans: ref_rev1
\begin{eqnarray}
\nonumber ds^2 = \alpha'\Big [ \frac{u^2}{R^2} \left ( (-V + F_0h^r_{tt}) dt^2 + dx_{||}^2 \right )\\ 
+ \frac{R^2}{u^2}\left ( (\frac{1}{V} + F_0h^r_{uu}) du^2 + (u^2+...) d\Omega_5^2 \right ) \Big ],
\end{eqnarray}
%gans: ref_rev1
with $F_0 = F_1 e^{-mz} \approx F_1(1 - mz)$.
The values of $g_{\alpha \beta} = G_{MN}\partial_{\alpha}X^M\partial_{\beta}X^N$, with $\tau = i\,t$, are:\\
\begin{math}
g_{00} = \frac{u^2}{R^2}V - \frac{u^2 F_0 h^r_{tt}}{R^2}; g_{11} = \frac{u^2}{R^2} + \frac{R^2}{u^2}(1/V + F_0 h^r_{uu})u'^2; 
\end{math}
\\
and 
\begin{math}
g_{01} = g_{10} = 0.
\end{math}

After ignoring the second order term $h^r_{tt}h^r_{uu}$, this gives 
\begin{eqnarray}
\label{eq:Spert}
%%%% gans: modification
%\nonumber S = \frac{1}{2\pi}\int d\sigma dt\\ 
\nonumber S = \frac{1}{2\pi}\int_{\Sigma} d\sigma d\tau\\ 
\nonumber \times \sqrt{\frac{u^4 V}{R^4} - \frac{u^4F_0 h^r_{tt}}{R^4} + (1 -  \frac{F_0 h^r_{tt}}{V} +F_0 Vh^r_{uu})u'^2 }.\\
\end{eqnarray}
%%%%%%%%%%%%%%%%%%%%%%%%%%%%%%%%%%%%%%%%%%
We note that the integrand is independent of $\tau$.
Apply the transformation: $\tau'(1 + \eta \sigma'/\beta) \leftarrow \tau$ and $\sigma' \leftarrow \sigma$, with $\eta = (\Delta \beta)/L$. The Jacobian of this transformation = J = $(1 + \eta \sigma/\beta)$.
Retaining $\sigma' = \sigma$, the integral then transforms as:
\begin{eqnarray}
\label{eq:sum_area}
\nonumber \int_{\Sigma} d\sigma d\tau (...) \rightarrow \int_0^L \int_0^\beta  J d\sigma d\tau'(...)\\
\nonumber = \int_0^L  \int_0^\beta d\sigma d\tau'(...) + \int_0^L \int_0^\beta  \frac{\eta\sigma}{\beta} d\sigma d\tau'(...)\\
= \int_0^L  \int_0^\beta d\sigma d\tau'(...) + \int_0^{\Delta \beta/L} d\theta' \int_0^L \sigma d\sigma(...),
\end{eqnarray}
where $\theta'$ is at this point, a dummy variable of integration.
%gans
The two integrals are easily seen to correspond to the rectangular and triangular regions ($\Sigma_1$ and $\Sigma_2$ respectively) in Fig.~\ref{fig:single_loop}. Figure~\ref{fig:strings} explicitly shows that the second integral $\int_0^{\Delta \beta/L}   d\theta' \int_0^L \sigma d\sigma(...)$ represents the area of the triangular region, with $\theta' = \theta$, and if $\Delta \beta$ is small. For a small value of $\Delta \beta$, the variables $\theta$ and $\sigma$ act as the "polar" coordinates for the triangular region.
We denote the first and second integrals in Eq.~\ref{eq:sum_area} as $S_1$ and $S_2$ respectively. 
The calculation of $S_1$ and $S_2$ is performed in the next two subsections, namely, Sec. \ref{sec:S1} and \ref{sec:S2}.

%gans:ref_rev1
Before proceeding with the calculation of the string action $S_1$ and $S_2$, we analyze qualitatively the effect of the shape of the contour $\Sigma$. The contour in Fig.~\ref{fig:strings_full} is a straight  line between $\beta$ and $\beta + \Delta \beta$. This means that $\beta$ varies linearly between $\sigma$ = $0$ and $L$. In actuality, it could be any curve $\beta(\sigma)$, based on the temperature variation of the thermal bath between the Quark and anti-Quark.
For the same value of $\beta$ and $\beta + \Delta \beta$, the worldsheet area given by the contour $\Sigma$ would differ for different curves $\beta(\sigma)$. 
Since $\Sigma$ forms the integration limits of Eq.~\ref{eq:Spert}, different $\Sigma$ could lead to different string action, S, in Eq.~\ref{eq:Spert}. This in turn could imply different values of $V_{Q\bar{Q}}$. This result seems expected since $V_{Q\bar{Q}}$ can depend not only on the temperature of the Quark and anti-Quark, but also on the temperature of the thermal medium between the two Quarks.
%gans:ref_rev1

\subsection{Calculation of Action $S_1$}
\label{sec:S1}
We now calculate the action $S_1$, i.e. the first integral in the last line of Eq.~\ref{eq:sum_area}.
%in Eq.~ref{eq:Spert}.
\begin{eqnarray}
\label{eq:Spert2}
%%%% gans: modification
%\nonumber S = \frac{1}{2\pi}\int d\sigma dt\\ 
\nonumber S_1 = \frac{1}{2\pi}\int_{\Sigma_1} d\sigma d\tau\\ 
\nonumber \times \sqrt{\frac{u^4 V}{R^4} - \frac{u^4F_0 h^r_{tt}}{R^4} + (1 -  \frac{F_0 h^r_{tt}}{V} +F_0 Vh^r_{uu})u'^2 }.\\
\end{eqnarray}
%%%%%%%%%%%%%%%%%%%%%%%%%%%%%%%%%%%%%%%%%%
The Hamiltonian, $H$, is then given by, 
\begin{equation}
H = \frac{\frac{u^4V}{R^4} - \frac{u^4 F_0 h^r_{tt}}{R^4} } {\sqrt{ \frac{u^4V}{R^4} + u'^2 - \left \{ \frac{u^4}{R^4} F_0 h^r_{tt} + F_0 D^ru'^2 \right \}  }},
\end{equation}
with  $D^r = \frac{h^r_{tt}}{V} - Vh^r_{uu}$, in terms of the redefined perturbation functions.
At minima, $u'=0$ and $u = u_0$, giving
\begin{equation}
H_0 = \frac{\frac{u_0^4V_0}{R^4} - \frac{u_0^4}{R^4}h^r_{tt}F_0}{\sqrt{ \frac{u_0^4V_0}{R^4} - \frac{u_4^2}{R^4}h^r_{tt}F_0}},
\end{equation}
with $V_0$ being $V$ evaluated at $u=u_0$.
The perturbation being small is unlikely to modify the string $u$ significantly. For the purpose of determination of the string equation, we ignore the perturbation to get,
\begin{equation}
\label{eq:h0}
H_0 = \frac{u_0^2\sqrt{V_0}}{R^2} ,
\end{equation}
which is a constant of motion. This gives,
\begin{equation}
%\nonumber \frac{u'}{u_0} = \frac{u_0}{R^2}\sqrt{\frac{u^4}{u_0^4} - \frac{a^4}{u_0^4} } \sqrt{\frac{u^4}{u_0^4} - 1}.
\nonumber \frac{u'}{a} = \frac{a}{R^2\sqrt{\frac{u_0^4}{a^4} - 1}}\sqrt{\frac{u^4}{a^4} - \frac{u_0^4}{a^4} }\sqrt{\frac{u^4}{a^4} - 1} .
\end{equation}

Again, defining $y$ = $u/a$, and $\omega$ = $u_0/a$,
\begin{equation}
\label{eq:ycurve}
y' = \frac{a}{R^2\sqrt{\omega^4 -1}} \sqrt{y^4 - \omega^4} \sqrt{y^4 - 1}.
\end{equation}
%gans
This gives:
\begin{eqnarray}
\nonumber a\int_{L/2}^L dr = aL/2\\ 
= R^2\sqrt{\omega^4 - 1}\int_{\omega}^{\infty} \frac{dy}{\sqrt{y^4 - \omega^4}\sqrt{y^4 - 1}},
\end{eqnarray}
or 
\begin{eqnarray}
\label{eq:lval}
\nonumber a = \frac{2 R^2\sqrt{\omega^4 - 1}}{L}\int_{\omega}^{\infty} \frac{dy}{\sqrt{y^4 - \omega^4}\sqrt{y^4 - 1}}\\
 = \frac{2 R^2\sqrt{\omega^4 - 1}}{L}I(R,\omega),
\end{eqnarray}
where (the integral $I(R,\omega)$ has been evaluated in~\cite{soo}):
\begin{description}
\item $I(R,\omega) = \int_{\omega}^{\infty} \frac{dy}{\sqrt{y^4 - \omega^4}\sqrt{y^4 - 1}}$\\
= $\frac{1}{4{\sqrt\gamma}{\sqrt(\omega^3)}}\Big [ K(\sqrt{\frac{\omega/2 + 1/(2\omega) + 1}{(\omega + 1/\omega)}}) \\
~~~~~~~~~- K(\sqrt{\frac{\omega/2 + 1/(2\omega) - 1}{\omega + 1/\omega}}) \Big ]$, with $K$ being the complete elliptic integral of the first kind;
\item $y = u/a$, 
\item $\omega$ = $u_0/a$, with $u_0$ being the minimum of u,
\item $\gamma = \frac{1}{2}(a + 1/a)$.
\end{description}
From this, one can determine $\omega$, and thus, $u_0$ implicitly.

We now look at the behavior of the string in Eq.~\ref{eq:ycurve} when $y$ (or equivalently $u$) $\rightarrow~\infty$. Taking the limit of large $y$, we get,
\begin{equation}
\label{eq:largey}
\lim_{y\rightarrow \infty} y' = \frac{a}{R^2\sqrt{\omega^4 -1}} y^4.
\end{equation}

%The action in Eq.~\ref{eq:Spert} can be written as the sum of two actions $S_1$ + $S2$
Coming back to the equation for action $S_1$, Eq.~\ref{eq:Spert2} can be rewritten as: 
\begin{eqnarray}
\nonumber S_1=\frac{1}{2\pi}\int_{\Sigma_1} d\tau d\sigma \sqrt{u'^2 + \frac{u^4V}{R^4}}\\
\times \left [ 1 + \frac{1}{2}\frac{\left ( -\frac{u^4}{R^4}h^r_{tt} - D^ru'^2\right )F_0 }{\frac{u^4V}{R^4} + u'^2} \right ].
\end{eqnarray}

The action $S_1$ is further split into two:
\begin{eqnarray}
S_{1a}= \frac{1}{2\pi} \int_{\Sigma_1} d\tau d\sigma \sqrt{u'^2 + \frac{u^4V}{R^4}},~~~~~~~~~\\
\nonumber S_{1b} = \frac{1}{2\pi} \int_{\Sigma_1} d\tau d\sigma\left [ \frac{1}{2}\frac{\left (-\frac{u^4}{R^4}h^r_{tt} - D^ru'^2\right ) F_0}{\frac{u^4V}{R^4} + u'^2} \right ].\\
\end{eqnarray}

The action $S_{1a}$, can be evaluated using standard techniques. The action $S_{1b}$ has to be evaluated numerically. 

\subsubsection{Calculation of $S_{1a}$}
Using standard techniques~\cite{mald1, mald2, soo}, and after subtracting the contribution of the self energy of the heavy Quark and anti-Quark system to the action, it can be seen that 

\begin{eqnarray}
\label{eq:s1}
\nonumber S_{1a}  = 2\beta \frac{a}{2\pi}\Big [ \int_\omega^{\infty} dy \left ( \frac{\sqrt{y^4 - 1}}{\sqrt{y^4 - \omega^4}} - 1 \right ) - \omega \Big ]\\ 
= 2\beta \frac{a}{2\pi}J(\omega),
\end{eqnarray}
where $J(\omega) = \int_\omega^{\infty} dy \left ( \frac{\sqrt{y^4 - 1}}{\sqrt{y^4 - \omega^4}} - 1 \right ) - \omega$.
%\begin{equation}
%\label{eq:s1}
% = \beta \frac{I(R,y,\omega)}{2\pi L} \left ( \frac{-(2\pi)^{3/2}}{\Gamma(1/4)^2} \right ).
%\end{equation}

\subsubsection{Convergence of $S_{1b}$}

While this has to be evaluated numerically, it is possible to confirm that it converges. 

\begin{eqnarray}
\label{eq:S1b}
\nonumber S_{1b} = \frac{1}{2\pi} \int_{\Sigma_1} d\tau d\sigma \sqrt{\frac{u^4V}{R^4} + u'^2}\\
\times \frac{1}{2}\Bigg ( \left [ \frac{-\frac{u^4}{R^4}h^r_{tt} }{\frac{u^4V}{R^4} + u'^2} \right ] - 
\left [ \frac{D^ru'^2}{\frac{u^4V}{R^4} + u'^2} \right ] \Bigg ) F_0,
\end{eqnarray}
with $D^r = \left ( \frac{h^r_{tt}}{V} - V h^r_{uu} \right )$ as mentioned earlier.

%%%%%%%%%%%%%%%%%%%%%%%%%%%%%%%%%%%%%%%%%%%%%%%%%%%%%%%%%%%%%
Let us take the behavior of $u' = ay'$, in the limit of large $u$ from Eq.~\ref{eq:largey}. From Fig.~\ref{fig:curvefit}, $\frac{h_{tt}}{V_G}$, and $V_G h_{uu}$, and hence $h^r_{tt}$, $h^r_{uu}$ and $D^r$ approximately behave as $u^{-2}$ in the limit of large $u$. 
We take $h^r_{tt} = A_{\infty}/u^{2}$, and $D^r = D_{\infty}/u^{2}$. We split the integral into two. One integral is till some finite, but large value of $u$ (= $u_{large}$), and thus evaluates to a finite result. The other integral is over the region covering $u_{large}$ to $\infty$. Thus, we get,  
\begin{eqnarray}
\nonumber S_{1b} \approx finite~value\\
\nonumber - \frac{1}{2\pi} \int_{\Sigma(u_{large})} d\tau d\sigma \sqrt{\frac{u^4V}{R^4} + u'^2}\\
\times \frac{1}{2}\Bigg ( \left [ \frac{A_{\infty }}{\frac{a^{-4}}{R^4(\omega^4 -1)}u^{6}} \right ] + 
\left [ \frac{D_{\infty}}{u^{2}} \right ] \Bigg ) F_0.
\end{eqnarray}
$F_0$ is bound by the value of $F_0$ at $z = 0$ = $F_{0max}$(say). Using this bound value, and expressing in terms of $y$, it gives,
\begin{eqnarray}
\nonumber |S_{1b}| < |finite~value| + |\frac{\beta a}{2\pi} \int_{y_{large}}^{\infty}  dy \frac{\sqrt{y^4 - 1}}{\sqrt{y^4 - \omega^4}}\\
\times \frac{1}{2}\Bigg ( \left [ \frac{A_{\infty} }{\frac{a^{2}}{R^4(\omega^4 -1)}y^{6}} \right ] + 
\left [ \frac{D_{\infty}}{a^{2}y^{2}} \right ] \Bigg ) F_{0max} |.
\end{eqnarray}
Given that $\lim_{y \rightarrow \infty} \frac{\sqrt{y^4 - 1}}{\sqrt{y^4 - \omega^4}} = 1$, and $\frac{\sqrt{y^4 - 1}}{\sqrt{y^4 - \omega^4}}$ is always bounded and $< J_{max}$ for $y > y_{large} (= \frac{u_{large}}{a})$, we get,
\begin{eqnarray}
\nonumber |S_{1b}| < |finite~value| + |\frac{\beta a }{2\pi}J_{max} \int_{y_{large}}^{\infty}  dy \\
\times \frac{1}{2}\Bigg ( \left [ \frac{A_{\infty} }{\frac{a^{2}}{R^4(\omega^4 -1)}y^{6}} \right ] + 
\left [ \frac{D_{\infty}}{a^{2}y^{2}} \right ] \Bigg ) F_{0max}|.
\end{eqnarray}
This integral clearly converges.
Hence, $S_{1b}$ seems to converge from the perspective of a numerical analysis, i.e., there exists at least some parameter values for which $S_{1b}$ converges.
The above analysis also indicates that as long as $D^r = -\frac{h^r_{tt}}{V} + Vh^r_{uu}$ or equivalently, $-\frac{h_{tt}}{V_G} + V_Gh_{uu}$, decay at a rate faster than $\frac{1}{u}$, for large $u$, $S_{1b}$ should converge. 
%%%%%%%%%%%%%%%%%%%%%%%%%%%%%%%%%%%%%%%%%%%%%%%%%%%%%%%%%%%%%

From Eq. \ref{eq:beta}, we have $E\propto \Delta \beta$. This implies $A$, $B$, and thus $h^r_{tt}$, $h^r_{uu}$ and $W^r$ are $\propto \Delta \beta$. Taking the implicit $\Delta \beta$ outside the expression for $S_{1b}$, we can write
\begin{equation}
\label{eq:finalS1b}
S_{1b} = \Delta \beta S^0_{1b}.
\end{equation}
%%%%%%%%%%%%%%%%%%%%%%%%%%%%%%%%%%%%%%%%%%%%%%%%%%

%gans
\subsection{Calculation of Action $S_2$}
\label{sec:S2}
\begin{eqnarray}
\nonumber S_2=\frac{1}{2\pi}\int_{\Sigma_2} d\tau d\sigma  \sqrt{u'^2 + \frac{u^4V}{R^4}}\\
\times \left [ 1 + \frac{1}{2}\frac{\left ( -\frac{u^4}{R^4}h^r_{tt} - 5W^ru'^2 \right )F_0 }{\frac{u^4V}{R^4} + u'^2} \right ].
\end{eqnarray}

In $S_2$, perturbation terms in the integrand lead to second order terms which involve the product of perturbations (as the domain $\Sigma_2$ already contains the perturbation to the metric). Thus ignoring the perturbation terms in the integrand for $S_2$,
\begin{equation}
S_2=\frac{1}{2\pi}\int_{\Sigma_2} d\tau d\sigma  \sqrt{u'^2 + \frac{u^4V}{R^4}}.
\end{equation}
%%%   As discussed, for calculation of the action $S2$, we continue with the same metric in Eq.~\ref{eq:base_metric} for both the particles.
%%%   The string action for the triangular region, with $\tau = \beta$ and $\sigma = x_{||}$, would be given by,
%%%   \begin{equation}
%%%   \nonumber S_2 = \frac{1}{2\pi} \int_{\Sigma_2} d\sigma d\tau \sqrt{u'^2 + \frac{u^4}{R^4}V},
%%%   \end{equation}
%%%   where $\Sigma_2$ represents the triangular region shown in Fig.~\ref{fig:strings}.
\begin{figure}[h!]
\includegraphics[width = 80mm,height = 80mm]{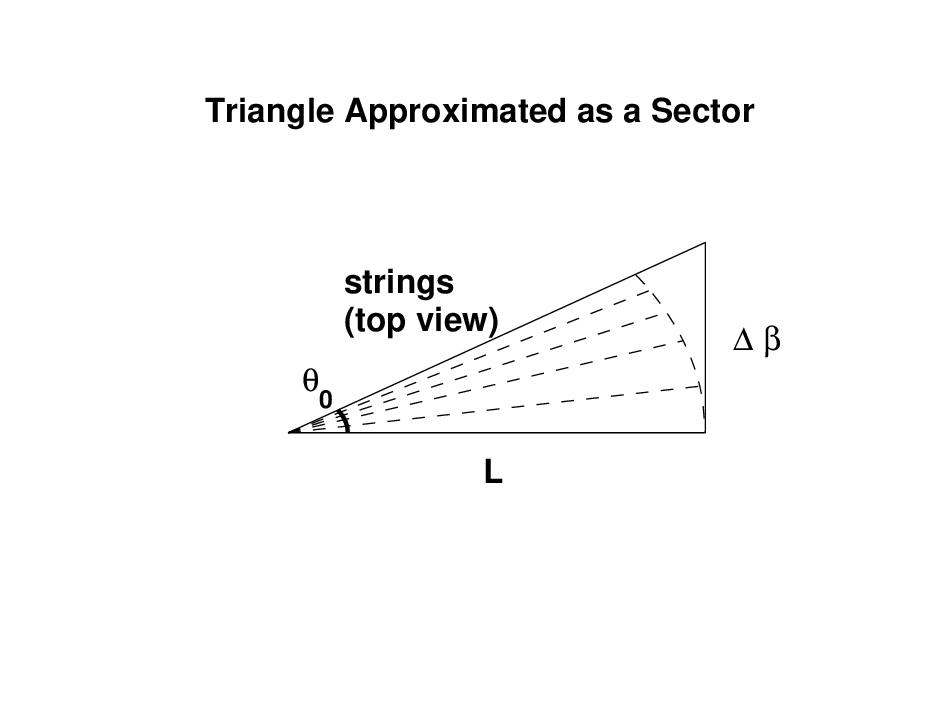}
\captionsetup{justification=raggedright, singlelinecheck=false}
\caption{Dashed lines represent the string contours in top view in the triangular region. The triangular region is approximated as a sector of a circle.}
\label{fig:strings}
\end{figure}
%gans
%end gans
The evaluation of the $S_2$ in terms of the polar coordinates had naturally risen from evaluation of the second integral in Eq.~\ref{eq:sum_area}.
Also, as can be seen in Fig.~\ref{fig:strings}, the string contours in $\Sigma_2$, i.e. the triangular region are most naturally represented in polar coordinates. In polar coordinates,
\begin{equation}
\label{eq:action}
S_2 = \frac{1}{2\pi}\int d\theta dr r \sqrt{u'^2 + \frac{u^4}{R^4}V},
\end{equation}
where, $u$ and $V$ are now expressed in polar coordinates. From the symmetry of the problem, $u$ would depend on $r$ alone.
The Lagrangian explicitly depends on $r$, and hence the Hamiltonian would also explicitly depend on $r$. The Hamiltonian, $H^a$ = 
\begin{equation}
\nonumber \frac{rVu^4/R^4}{\sqrt {u'^2 + Vu^4/R^4 }}.
\end{equation}
We express the above Hamiltonian as $H^a = r H^a_0$. We have separated out the constant $H^a_0$ from the $r$ dependent Hamiltonian.
If $u_0$ is the value of $u$ at the minima (i.e. at $u'$ =0), then $H^a_0 = \frac{u_0^2\sqrt{V_0}}{R^2}$. The value of the Hamiltonian density $H^a_0$ is the same as that of the Hamiltonian density, $H_0$, we obtained in Eq.~\ref{eq:h0}. 
%gans
This equality is required, and additionally, acts as a check, as the two Hamiltonian densities belong to different regions of the same Polyakov loop correlator, describing a single (same) Quark anti-Quark system. 
Since the different regions are parts of the same system, and since $H_0$(Eq.~\ref{eq:h0}) is a constant of motion, we infer that $H^a_0$ is also a constant of motion.
This gives
\begin{equation}
\nonumber r H^a_0 = \frac{r Vu^4/R^4 }{\sqrt{u'^2 + Vu^4/R^4 }}.
\end{equation}
Substituting $H^a_0$ and solving for $u$,
\begin{equation}
\nonumber \frac{u'}{a} = \frac{a}{R^2\sqrt{\frac{u_0^4}{a^4} - 1}}\sqrt{\frac{u^4}{a^4} - \frac{u_0^4}{a^4} }\sqrt{\frac{u^4}{a^4} - 1} .
%\nonumber \frac{u'}{u_0} = \frac{u_0}{R^2}\sqrt{\frac{u^4}{u_0^4} - \frac{a^4}{u_0^4} } \sqrt{\frac{u^4}{u_0^4} - 1}.
\end{equation}
It is seen that the explicit $r$ dependence is eliminated.
Again, defining $y$ = $u/a$, and $\omega$ = $u_0/a$,
\begin{equation}
\nonumber y' = \frac{a}{R^2\sqrt{\omega^4 -1}} \sqrt{y^4 - \omega^4} \sqrt{y^4 - 1}.
%\nonumber y' = \frac{u_0}{R^2} \sqrt{y^4 - \omega^4} \sqrt{y^4 - 1}.
\end{equation}
%gans
%gans
This gives:
%\begin{equation}
%\nonumber \int dr = \frac{R^2}{U_0}\int_{\omega}^{U/U_0} \frac{dy}{\sqrt{y^4 - \omega^4}\sqrt{y^4 - 1} }.
%\end{equation}
%From, symmetry of the problem,
\begin{eqnarray}
\label{eq:rval}
\nonumber a\int_{0}^{L/2} dr = aL/2\\
 = R^2\sqrt{\omega^4 - 1}\int_{\omega}^{\infty} \frac{dy}{\sqrt{y^4 - \omega^4}\sqrt{y^4 - 1}},
%\int_{L/2}^L dr = L/2 = \frac{R^2}{u_0}\int_{\omega}^{\infty} \frac{dy}{\sqrt{y^4 - \omega^4}\sqrt{y^4 - 1}},
\end{eqnarray}
or 
\begin{eqnarray}
\label{eq:a2}
%u_0 = \frac{2 R^2}{L}\int_{\omega}^{\infty} \frac{dy}{\sqrt{y^4 - \omega^4}\sqrt{y^4 - 1}} = I(R,y,\omega)/L,
\nonumber a = \frac{2 R^2\sqrt{\omega^4 - 1}}{L}\int_{\omega}^{\infty} \frac{dy}{\sqrt{y^4 - \omega^4}\sqrt{y^4 - 1}}\\
 = \frac{2 R^2\sqrt{\omega^4 - 1}}{L}I(R,\omega),
\end{eqnarray}
with $I(R,\omega)$ as defined earlier.

Substituting the expression for $y'$ and $dr$ in the action in Eq.~\ref{eq:action}, and noting that the integrand is independent of $\theta$, we get the result, 
\begin{eqnarray}
\nonumber S_{2}  = \theta_0 \frac{a}{2\pi} \int_\omega^{\infty} r dy \left ( \frac{\sqrt{y^4 - 1}}{\sqrt{y^4 - \omega^4}}  \right )\\ 
\nonumber   - \theta_0 \frac{a}{2\pi} \int^\omega_{\infty} r dy \left ( \frac{\sqrt{y^4 - 1}}{\sqrt{y^4 - \omega^4}}  \right ).
%\nonumber S_2  = \frac{u_0}{2\pi} \theta_0 \Big ( \int_1^{u/u_0} dy\,r(y) \frac{y^2}{\sqrt{y^4 - 1}} \\
%\nonumber + \int^1_{u/u_0} dy\,r(y) \frac{y^2}{\sqrt{y^4 - 1}} \Big ).
\end{eqnarray}
The string extends from $\infty$ to $u_0$ and then from $u_0$ to $\infty$. Hence, the integration is divided into two limits, $\omega$ to $u/a$ and $u/a$ to $\omega$. The heavy Quark anti-Quark system self energy needs to be subtracted. The self energy contribution to the action (and remembering that $u$ is now in polar coordinates) = $S_{SE}$
%$~\equiv~\beta \times self~energy~$ \\ 
$\equiv~\frac{1}{2\pi}\int_0^{\theta_0} r d\theta \int_{u_0}^u du - \frac{1}{2\pi}\int_0^{\theta_0} r d\theta \int^{u_0}_u du+ L\theta_0\frac{u_0}{2\pi}$ \\
= $a \theta_0\int_{\omega}^{u/a}r dy - a \theta_0\int^{\omega}_{u/a}r dy+  L\theta_0\frac{a\omega}{2\pi}$.\\
After subtracting $S_{SE}$ from $S_2$, $S_2$ becomes,
\begin{eqnarray}
\nonumber S_{2}  = \theta_0 \frac{a}{2\pi } \int_\omega^{\infty} r dy \left ( \frac{\sqrt{y^4 - 1}}{\sqrt{y^4 - \omega^4}} -1 \right )\\ 
\nonumber   - \theta_0 \frac{a}{2\pi } \int^\omega_{\infty} r dy \left ( \frac{\sqrt{y^4 - 1}}{\sqrt{y^4 - \omega^4}} -1 \right ) - L\theta_0\frac{a\omega}{2\pi}.
%\nonumber S_2  = \frac{u_0}{2\pi} \theta_0 \Bigg ( \int_1^{u/u_0} dy\,r(y) \left ( \frac{y^2}{\sqrt{y^4 - 1}} -1 \right ) \\
%\nonumber + \int^1_{u/u_0} dy\,r(y) \left ( \frac{y^2}{\sqrt{y^4 - 1}} -1 \right )  - L\frac{u_0}{2\pi} \Bigg ).
\end{eqnarray}
As $y$ ranges from $u/a$ to $\omega$, $r$ ranges from $0$ to $L/2$, and as  
$y$ ranges from $\omega$ to $u/a$, $r$ ranges from $L/2$ to $L$.
Recognizing that for small $\theta_0$, $\theta_0$ = $\frac{\Delta \beta}{L}$,
\begin{eqnarray}
\nonumber S_{2}  = \frac{\Delta \beta}{L} \frac{a}{2\pi} \int_\omega^{\infty} r dy \left ( \frac{\sqrt{y^4 - 1}}{\sqrt{y^4 - \omega^4}} -1 \right )\\ 
\nonumber - \frac{\Delta \beta}{L} \frac{a}{2\pi} \int^\omega_{\infty} r dy \left ( \frac{\sqrt{y^4 - 1}}{\sqrt{y^4 - \omega^4}} -1 \right ) - L\frac{\Delta \beta}{L}\frac{a \omega}{2\pi}.\\ 
%\nonumber S_2  = \frac{I(R,y,\omega)}{2\pi L} \frac{\Delta \beta}{L} \Bigg \{ \int_1^{u/u_0} dy\,r(y) \left ( \frac{y^2}{\sqrt{y^4 - 1}} -1 \right ) \\
%\nonumber + \int^1_{u/u_0} dy\,r(y) \left ( \frac{y^2}{\sqrt{y^4 - 1}} -1 \right )   - L\frac{u_0}{2\pi} \Bigg \}.
\end{eqnarray}
Integrating by parts,
\begin{eqnarray}
\nonumber S_2  = \frac{a \Delta \beta}{2\pi L} ~~~~~~~~~~~~~~~~~~~~\\
%\nonumber \times \left ( \Big \{ L\int_1^{\infty} dy\,\left ( \frac{y^2}{\sqrt{y^4 - 1}} -1 \right ) \Big \} -L \right ) \\ 
\nonumber \times \Bigg \{ \Big [ r \int dy\,\left ( \frac{\sqrt{y^4 - 1}}{\sqrt{y^4 - \omega^4}} -1 \right ) \Big ]_{y=\omega}^{y=\infty} \\
\nonumber -  \Big [ r \int dy\,\left ( \frac{\sqrt{y^4 - 1}}{\sqrt{y^4 - \omega^4}} -1 \right ) \Big ]_{y=\infty}^{y=\omega} 
-L\omega \Bigg \} \\ 
\nonumber - \frac{a \Delta \beta}{2\pi L} 2 \int_{0}^{L/2} dr \int dy \left ( \frac{\sqrt{y^4 - 1}}{\sqrt{y^4 - \omega^4}} -1 \right),
\end{eqnarray}
with $dr$ given by Eq.~\ref{eq:rval}. This then gives,
%\nonumber - \frac{a \Delta \beta}{2\pi L} 2 \int_{\omega}^{\infty} dy \int dy \left ( \frac{\sqrt{y^4 - 1}}{\sqrt{y^4 - \omega^4}} -1 \right).
%Solving the integrals,
\begin{eqnarray}
\label{eq:s2}
S_2  = a \Delta \beta \left \{ \frac{1}{2\pi} J(\omega)  
- \frac{C_0}{2\pi L}  \right \},
\end{eqnarray}
with $C_0$ being a positive constant and is equal to~ 
\begin{math}
%2 \int_{\omega}^{\infty} dy \int dy \left ( \frac{\sqrt{y^4 - 1}}{\sqrt{y^4 - \omega^4}} -1 \right ).
2 \int_{0}^{L/2} dr \int dy \left ( \frac{\sqrt{y^4 - 1}}{\sqrt{y^4 - \omega^4}} -1 \right ).
\end{math}
%To see the explicit dependence on L, one can substitute $a$ from Eq.\ref{eq:a2}, to obtain,
%\begin{eqnarray}
%S_2  = \frac{2 R^2\sqrt{\omega^4 - 1}}{L}I(R,\omega) \Delta \beta \left \{ \frac{1}{2\pi} J(\omega)  
%- \frac{C_0}{2\pi L}  \right \}.
%\end{eqnarray}

\subsection{Calculation of inter Quark potential $V_{Q\bar{Q}}$}
From Eqs.~\ref{eq:s1}, \ref{eq:finalS1b} and \ref{eq:s2}, and taking $\beta \sim 1/\Theta$, the final potential between the heavy Quark and anti-Quark is: 
%gans
\begin{eqnarray}
\label{eq:finalV}
\nonumber \langle V_{Q\bar{Q}}(L,T)\rangle =  S/\beta \sim S\Theta~~~~~~~~~~~~~~~~~~~~\\
\nonumber =  a \Big [ \frac{1}{2\pi} \left \{ (2 + \Theta \Delta \beta) \left ( J(\omega) \right )  \right \} \\
- \frac{\Theta \Delta \beta}{2\pi L}C_0 \Big ] + \Theta \Delta \beta S^0_{1b}.
\end{eqnarray}
%\nonumber =  I(R,y,\omega) \Big [ \frac{1}{2\pi L} \left \{ (2 + \frac{\Delta \beta}{\beta}) \left ( \frac{-(2\pi)^{3/2}}{\Gamma(1/4)^2} \right )  \right \} \\
%- \frac{\Delta \beta/\beta}{2\pi L^2}C \Big ].
%\nonumber =  a \Big [ \frac{1}{2\pi} \left \{ (2 - \Theta \Delta \beta + \Theta \Delta \beta) \left ( J(\omega) \right )  \right \} \\
%- \frac{\Theta \Delta \beta}{2\pi L}C_0 \Big ] + \Theta \Delta \beta S^0_{1b}.
%gans
In terms of the linear perturbation parameters of the black string metric, from Eq.~\ref{eq:beta}, 
%i.e., with $\beta=\pi a$ and $\Delta \beta = \pi aEmL/2$, 
it becomes,
\begin{eqnarray}
\label{eq:finalVper}
%\nonumber \langle V_{Q\bar{Q}}(L,T)\rangle =  ~~~~~~~~~~~~~~~~~~~~~~~~~~~~~~~~\\
\nonumber \langle V_{Q\bar{Q}}(L,T)\rangle =  a \Big [ \frac{1}{2\pi} \left \{ (2 + \frac{EmL}{2} ) \left ( J(\omega) \right )  \right \} \\
- \frac{Em}{4\pi}C_0 \Big ] + \frac{EmL}{2}S^0_{1b}.
%- \frac{\frac{EmL}{2}}{2\pi L^2}C_0 \Big ] + \frac{EmL}{2\pi a}S^0_{1b}.
\end{eqnarray}
%where its understood that heavy Quark self energy is subtracted from $S^0_{1b}$.

\begin{figure}[h!]
\includegraphics[width = 80mm,height = 80mm]{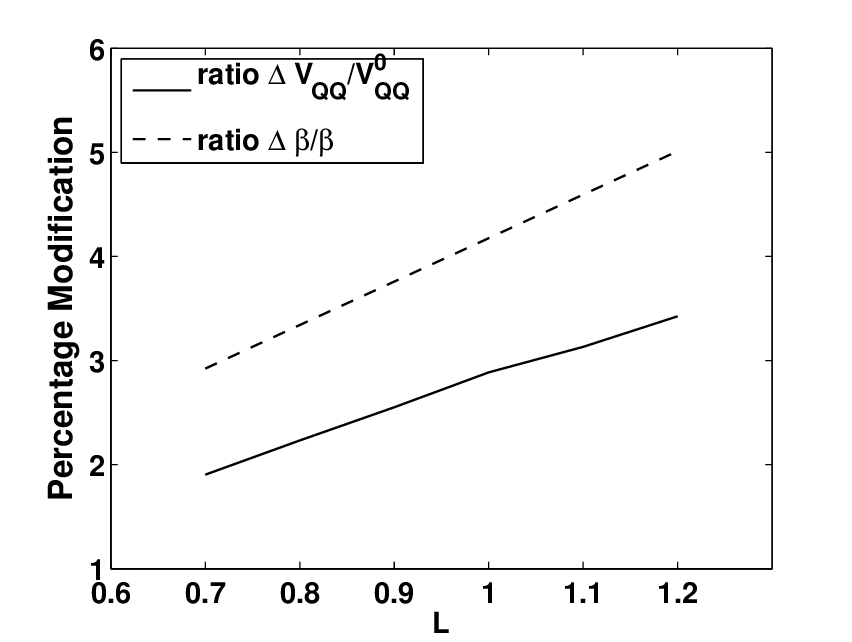}
\captionsetup{justification=raggedright, singlelinecheck=false}
\caption{Plot of ratios $\frac{\Delta V_{Q\bar{Q}}}{V^0_{Q\bar{Q}}}$ = $\frac{(S_{1b} + S_2)/\beta}{S_{1a}/\beta}$ and $\frac{\Delta \beta}{\beta}$.}
\label{fig:deltaV}
\end{figure}

The zero temperature gradient case is obtained by simply assigning $\Delta \beta = 0$. This gives,
\begin{eqnarray}
\label{eq:zeroV}
\langle V^0_{Q\bar{Q}}(L,T)\rangle =  
 a \Big [ \frac{1}{\pi L} J(\omega) \Big ].
\end{eqnarray}
The change $\Delta V_{Q\bar{Q}}$ is given by,
\begin{eqnarray}
\label{eq:deltaV}
%\nonumber \Delta \langle V_{Q\bar{Q}}(L,T)\rangle~~~~~~~~~~~~~~~~~~~~~~~~~~~\\
%\nonumber =  (S_{1b} + S_2)/\beta \sim (S_{1b} + S_2)\Theta~~~~~~~~~~~~~~~~~~~~~\\
\nonumber \Delta \langle V_{Q\bar{Q}}(L,T)\rangle =  (S_{1b} + S_2)/\beta = (S_{1b} + S_2)\Theta~~~~\\
\nonumber =  a \Big [ \frac{1}{2\pi} \left \{ (\Theta \Delta \beta) \left ( J(\omega) \right )  \right \} 
- \frac{\Theta \Delta \beta}{2\pi L}C_0 \Big ] + \Theta \Delta \beta S^0_{1b}.\\
\end{eqnarray}
Figure~\ref{fig:deltaV}, plots the ratio $\frac{\Delta V_{Q\bar{Q}}}{V^0_{Q\bar{Q}}}$ = $\frac{(S_{1b} + S_2)/\beta}{S_{1a}/\beta}$, for various values of L.  This plot uses the same perturbation parameters used for Figs.~\ref{fig:hall} and \ref{fig:curvefit}. It can be seen that this ratio is monotonically increasing. This is due to the fact that as $L$ increases, $\Delta \beta = \theta_0 L$ increases. The corresponding ratio $\frac{\Delta \beta}{\beta}$ is also shown alongside.  The monotonically increasing nature of $\frac{\Delta \beta}{\beta}$ reflects the fact that, as the inter-Quark distance increases, the temperature difference between the Quarks would also increase. The (almost) straight line for $\frac{\Delta V_{Q\bar{Q}}}{V^0_{Q\bar{Q}}}$, is a consequence of our linear approximation. 
%The value of $\beta = \frac{\pi R^2}{2a}$.

 As $\Delta \beta$ increases, i.e., the temperature decreases, for one of the Quarks, it is natural to expect the binding energy of $Q\bar{Q}$ to increase. This is qualitatively, what is seen in this analysis.

\subsection{Calculation of $d_{max}$}
\begin{figure}[h!]
\includegraphics[width = 80mm,height = 80mm]{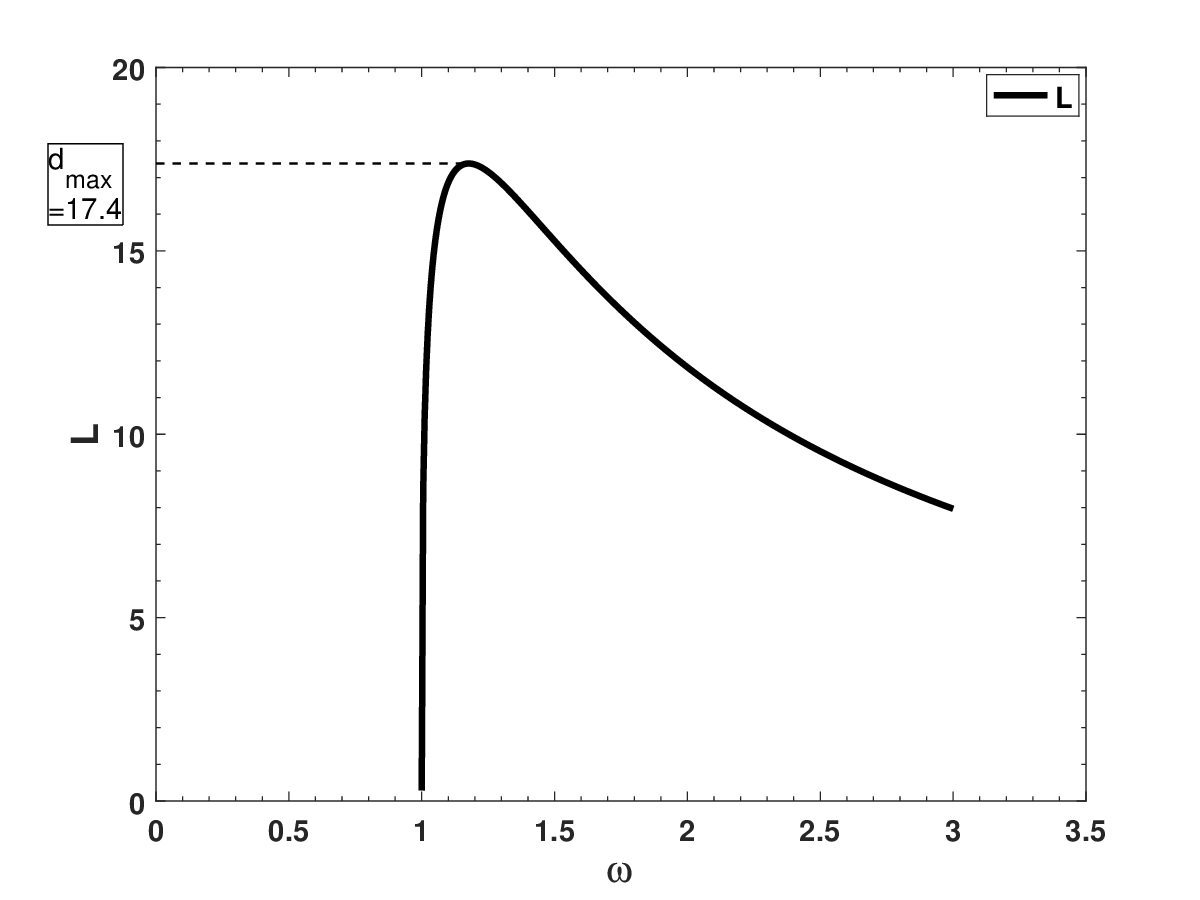}
\captionsetup{justification=raggedright, singlelinecheck=false}
\caption{Plot of inter-Quark distance L as a function of $\omega$.}
\label{fig:dmax}
\end{figure}

We now calculate the value of the maximum Quark anti-Quark separation, i.e., $d_{max}$. From Eq.~\ref{eq:lval}, one can determine the inter-Quark distance, $L$, as 
\begin{eqnarray}
\label{eq:dmax_val}
\nonumber L = \frac{2 R^2}{a}\sqrt{\omega^4 - 1}\,I(R,\omega).
\end{eqnarray}
Figure~\ref{fig:dmax} plots the inter-Quark distance, $L$, as a function of $\omega$ ($= \frac{u_0}{a}$).
It can be inferred from the plot, that the value of $d_{max}=17.4$ at $\omega = \omega_{max} = 1.17$. 
The values of $\omega$ used for simulation in Fig.~\ref{fig:deltaV} are greater than $\omega_{max}$. 
%It can be recalled that $\omega = \frac{u_0}{a}$ or $a\omega - u_0$. 
In Ref.~\cite{soo}, it is argued that the string corresponding to a  larger value of $\omega$ would be shorter and hence energetically, a more favorable configuration. 
\section{Summary and Conclusions}
\label{sec:conclusion}
The effect of a temperature gradient on the heavy Quark, anti-Quark potential has been analyzed and calculated. The temperature gradient in the dual AdS space is obtained by inserting a non-uniform black string. The non-uniform black string metric is developed by perturbing the Schwarzchild solution. The modification in the potential due to the temperature gradient is seen to be proportional to $\Delta \beta$ for small $\Delta \beta$. 
In our calculations, we have seen a small correction to the potential, which increases with increasing separation between the Quark and anti-Quark.
%The term contains a conformally invariant $L^{-1}$ term and a non-invariant $L^{-2}$ term. 
%The $L^{-2}$ term needs to be probed further, which we plan to do in future. 
%This term is not conformally invariant. A non conformally invariant term is a possibility at finite temperatures, since conformal invariance is broken at finite temperatures. 
This calculation, provides the heavy Quark anti-Quark potential, for a {\fontfamily{pzc}\selectfont N} $= 4$ supersymmetric Yang-Mills Lagrangian, instead of the QCD Lagrangian.
	
	Equation~\ref{eq:nonlintemp} provides a mechanism to model non-linear temperature gradients, though we believe that the subsequent string action would then have to be evaluated completely numerically. This could help in modeling larger temperature variations if present in any system.
%Quarkonium yield is used as a signature to study the presence and properties of QGP.
As a side note, recent literature models QGP using 2+1 or 3+1 hydrodynamics~\cite{2p1a, 3p1a}, which could make available, the temperature distribution inside the QGP.

%	While the results in this work may not be directly be applicable to a realistic QCD Lagrangian, 
The metric perturbation developed in this work, and suitably modified for four dimensional spacetime, might be used to compute QCD Polyakov correlators or QCD thermal propagators with temperature gradients for QCD systems in the perturbative regime.
We have touched upon this aspect in Sec.~\ref{sec:polyakov}, where we interpreted the nonuniform topological cylinder to be a uniform cylinder in perturbed Euclidean spacetime. We plan to pursue these aspects in future work. 

  It would also be useful to incorporate the proposed prescription within the framework of Schwinger Keldysh formulation.

\appendix
\section{Quasi-Stationarity}
\label{sec:quasi}
We analyze the quasi-stationarity in both the primal gauge theory and the dual AdS domain. 
\subsection{Primal Domain}
	The relation between viscosity and the mean free path is given by~\cite{book1, book2}:
\begin{equation}
%	\lambda = \frac{\mu}{p}\sqrt{\frac{\pi T}{2M}},
	\eta \propto \lambda,
\end{equation}
where $\eta$ and $\lambda$ are shear viscosity and mean free path, respectively.
%where $\lambda$ is the mean free path, $\mu$ is the viscosity, $p$, $T$ and $M$ are the pressure, temperature and mass respectively. 
%mass of what?
The QGP fluid has been experimentally seen to be an ideal fluid~\cite{naturephy}, in that the viscosity is negligible. Several other studies have also indicated the QGP to be a low viscosity, near perfect fluid~\cite{bk2ref52, bk2ref53, bk2ref54, bk2ref51}. Lattice calculations point in the same direction~\cite{etalat}. This means that the time of mean free path of the gluons and quarks inside the QGP is very small. 
Since the time of mean free path is very small, the QGP equilibrates and remains in local thermodynamic equilibrium even as the temperature keeps changing. Thus, the equilibration happens much faster than the time variation in temperature. In other words, the QGP medium can be treated as a quasi-stationary system. 
In fact, in literature, the temperature dependent quark antiquark potential, has been determined at constant temperature~\cite{Laine1}. This temperature dependent potential has then been used to solve a time independent Schroedinger equation, in a plasma whose temperature is varying with time ~\cite{Wolschin, vitev1, vitev2, Strickland1, Strickland2, book2}.
Solving the time independent Schroedinger equation determines the instantaneous Quarkonium wavefunction as a quasi-stationary approximation.
If the primal domain can be treated as a quasi-stationary system, then the dual domain is also expected to behave as a quasi-stationary system.  
There is however a subtlety here. The AdS domain is dual to a {\fontfamily{pzc}\selectfont N} $= 4$ supersymmetric SU(N) plasma, but the QGP is an SU(3) QCD plasma. Even in the supersymmetric SU(N) plasma, $\frac{\eta}{s}$ has been shown to be equal to a low value of $\frac{1}{4\pi}$~\cite{kovtun}. Thus local thermodynamic equilibrium is expected even in the SU(N) plasma and a quasi-stationary approximation should be reasonably accurate. 
Since the dual AdS domain models the same physical system, it should be possible to treat the dual AdS domain also in a quasi-stationary way.
\subsection{Dual domain}
	If spacetime is curved, the vacuum state behaves as if it contains particles, compared to another vacuum in flat spacetime, leading to Hawking radiation.
   If the spacetime curvature were to vary with time, $t$, at a rate $\propto~\kappa$, then it leads to a possibility that the probability of particles produced with energy $\frac{\kappa}{\sqrt{-g_{00}}}$ is altered. 
A particle of energy $\frac{\kappa}{\sqrt{-g_{00}}}$ would be produced over a time interval $\frac{1}{\kappa}$. But within this time interval, the curvature would have changed. Thus particle production probability (of energy $\frac{\kappa}{\sqrt{-g_{00}}}$), can get altered, and thus altering the Hawking radiation/temperature.
 If this modification of particle production is significant, then quasi-stationary approximation will not be valid.

We now calculate the Ricci scalar curvature and determine its rate of time variation, $\kappa$. Let us recall that the time dependency of the metric perturbation  = $e^{\Omega t}$, where $\Omega = -\frac{4a}{R^2}$. We are interested in the case of small $\Omega$ or equivalently, large $R$. This is also consistent with the large $N$ limit. 
The time dependent space time metric used for temperature calculation (from Eq.~\ref{eq:temp_metric} and noting that $V_G \approx \frac{4au_+}{R^2}$ near the horizon) is:
%get the right metric. Below is a place holder
\begin{eqnarray}
\nonumber ds^2 = \left (-\frac{4au_+}{R^2} + e^{\Omega t}e^{-mz}A u_+\right ) dt^2 \\
\nonumber +  \left ( \frac{R^2}{4au_+}+\frac{e^{\Omega t}e^{-mz}B}{u_+} \right ) du_+^2 \\
\nonumber + e^{\Omega t}e^{-mz} C du_{+}dt\\
+ R^2d\Omega_5^2 + Gdz^2.
\end{eqnarray}
%\begin{eqnarray}
%\nonumber ds^2 = \left (-V_G + e^{\Omega t'}e^{-mz}A u'_+\right ) dt'^2 \\
%\nonumber + \Big ( \frac{1}{V_G}+\frac{e^{\Omega t'}e^{-mz}B}{u'_+} \\
%\nonumber - \frac{e^{2\Omega t'}e^{-2mz} C^2a}{4(-V_G + e^{\Omega t'}Aae^{-mz}u_+)} \Big ) du'^2_{+}\\
%+ (...)d\Omega_5^2 + Gdz^2.
%\end{eqnarray}
In the limit of large $R$, the Ricci scalar curvature, $R_{T}$, for the above metric is given by:
\begin{equation}
	\lim_{R\rightarrow large}R_{T} = -\frac{R^2 m^2}{2u_+^2}.
\end{equation}
We see that the Ricci curvature, in the limit of large $R$, is independent of time! All the factors involving time cancel out, giving $\kappa \sim \frac{1}{R_T}(\nabla_t R_T) = \frac{1}{R_T}\frac{\partial R_T}{\partial t} \approx 0$.
Thus, to a first order, the modification of the vacuum production of particles, because of the time variation of the given curved spacetime metric should be subdued.
This again indicates that a quasi-stationary approach may be followed.
In this universe of large $R$, an observer at $t = t_0$, sees that the Ricci scalar curvature is constant, and can likely go about measuring the temperature based on the instantaneous metric present at $t=t_0$.

% du_+dt term is present. This means that horizon is changing with time. Will this lead to rate of change of curvature? Ans=no. Why? Note: u_+ is not a spatial axis. It is the energy axis, and related to temperature. The quark temperature is changing (along with the horizon). This means that as horizon changes, quark temperature also changes. This means du_+ continues to be zero. Hence dtdu_+ does not play a role.

\begin{thebibliography}{99}
\bibitem{nature}  P. Braun-Munzinger and Johanna Stachel, Nature, {\bf 448}, 302-309 (2007). %"The quest for the quark-gluon plasma"

\bibitem{naturephy}  PHENIX Collaboration,  Nature Physics, {\bf 12}, (2018); arxiv:nucl-ex/1805.02973 (2018).
\bibitem{PRCver}  PHENIX Collaboration, Phys. Rev. C, {\bf 97}, 064904 (2018); arXiv:nucl-ex/1710.09736 (2018). 
%\bibitem{viscous} Heinz, U., Song, H. and Chaudhuri, A. K.  Phys. Rev. C 73, 034904 (2006).  %Dissipative hydrodynamics for viscous relativistic fluids. 
\bibitem{mats} T. Matsui and H. Satz, Phys. Lett. B, {\bf 178}, 416 (1986). 
\bibitem{Chu} M. C. Chu and T. Matsui, Phys. Rev. D, {\bf 37}, 1851 (1988).
\bibitem{PKSMCA} M. C. Abreu et al. (NA50 Collaboration), Phys. Lett. B, {\bf 477}, 28 (2000); B. Alessandro et al. (NA50 Collaboration), Eur. Phys. J. C, {\bf 39}, 335 (2005).

\bibitem{PKSRAR} R. Arnaldi, et al. (NA60 Collaboration), Phys. Rev. Lett., {\bf 99}, 132302 (2007); R. Arnaldi (NA60 Collaboration), Presentation at the ECT workshop on "Heavy Quarkonia Production in Heavy-Ion Collisions," Trent (Italy), May 25-29 (2009).

\bibitem{PKSAAD} A. Adare et al. (PHENIX Collaboration), Phys. Rev. Lett., {\bf 98}, 232301 (2007).

\bibitem{PKSCMS1} The CMS Collaboration, JHEP, {\bf 05}, 063 (2012).

\bibitem{PKSBAB} B. Abelev et al. (ALICE Collaboration), Phys. Rev. Lett., {\bf 109}, 072301 (2012).
\bibitem{Madhu1} M. Mishra, C. P. Singh, V. J. Menon and Ritesh Kumar Dubey, Phys. Lett. B, {\bf 656}, 45 (2007).
\bibitem{gans1} S. Ganesh and M. Mishra, Phys. Rev. C, {\bf 88}, 044908 (2013).
\bibitem{gans2} S. Ganesh and M. Mishra, Phys. Rev. C, {\bf 91}, 034901 (2015).
%\bibitem{gans} S. Ganesh, M. Mishra, Phys. Rev. C {\bf 88}, 044908 (2013).
\bibitem{gans3} S. Ganesh and M. Mishra, Nuclear Phys. A., {\bf 947}, 38-63 (2016).
%\bibitem{mald1} J. Maldacena, "arXiv:hep-th/9803002v3", (1998)
\bibitem{mald1} J. M. Maldacena, Phys. Rev. Lett., {\bf 80}, 4859-4862 (1998); arXiv:hep-th/9803002v3 (1998).
%\bibitem{mald2} J. Maldacena, "arXiv:hep-th/9711200v3", (1998)
\bibitem{mald2} J. M. Maldacena, 
%"The large N limit of superconformal field theories and supergravity,"
Adv. Theor. Math. Phys. 2 231 (1998); Int. J. Theor. Phys., {\bf 38}, 1113 (1999);
arXiv:hep-th/9711200 (1998).
%%%%%%%%%%%%%%%%%%%%%
%viscosity paper:QGP
\bibitem{kovtun} P. K. Kovtun, D. T. Son and A.O. Starinets, Phys. Rev. Lett., {\bf 94}, 111601 (2005).
\bibitem{hotwind} Hong Liu, Krishna Rajagopal and Urs Achim Wiedemann, Phys. Rev. Lett., {\bf 98}, 182301 (2007); arxiv:hep-ph/0607062 (2006).
\bibitem{tatsuo2}       Tomoya Hayata, Kanabu Nawa and Tatsuo Hatsuda,
%"Time-dependent heavy quark potential at finite temperature from gauge-gravity duality",
Phys. Rev. D, {\bf 87}, 101901 (2013); arxiv:hep-ph/1211.4942 (2013).
%%%%%%%%%%%%%%%%%%%%%
%non-QGP
\bibitem{tatsuo1}       Tatsuo Hatsuda,
%"Heavy Quarkonium in Hot Medium",
%The Quark Matter 2012 Proceedings of the XXIII International Conference on Ultrarelativistic Nucleus-Nucleus Collisions,
Nucl. Phys. A, 904-905 (2013).
\bibitem{koji} Koji Hashimoto, Noriaki Ogawa and Yasuhiro Yamaguchi,
%"Holographic Heavy Quark Symmetry",
arxiv:hep-th/1412.5590v3 (2015).
\bibitem{hyo} Hyo Chul Ahn, Deog Ki Hong, Cheonsoo Park and Sanjay Siwach,
%Spin 3/2 Baryons and Form Factors in AdS/QCD, 
Phys. Rev. D, {\bf 80}, 054001 (2009); arxiv:hep-ph/0904.3731 (2009).

%%%%%%%%%%%%%%%%%%%%%
%\bibitem{introwilson} Yuri Makeenko, arxiv:hep-th/0906.4487v1 (2009).
\bibitem{soo} Soo-Jong Rey, Stefan Theisen and Jung-Tay Yee, Nuclear Phys. B, {\bf 527}, 171-186 (1998).
\bibitem{witten} E. Witten, Adv. Theor. Math Physics, 2:505-532 (1998).
%\bibitem{finite2} Kiyoshi Shiraishi, Progress of Theoretical Physics 78, No. 3  arxiv:hep-th/1408.5675v1 (2014).
%nonuniform black string
\bibitem{bs} Liu Zhao, Kai Niu, Bing-Shu Xia, Yi-Ling Dou and Jie Ren, Class. Quant. Grav., {\bf 24}, 4587-4600 (2007); arxiv:hep-th/0703195 (2007).
\bibitem{greg} R. Gregory and R. Laflamme, Phys. Rev. Lett., {\bf 70}, 2837-2840 (1993); arxiv:hep-th/9301052 (1993).
\bibitem{greg_book} R. Gregory, arxiv:1107.5821v1 [gr-qc] (2011).

%gans: new references
\bibitem{2p1a} Yunpeng Liu, Che Ming Ko, Taesoo Song, Physics Lett. B, {\bf 728}, 437 (2014).
\bibitem{3p1a} Li et. al., Phys. Rev. C, {\bf 98}, 014909 (2018).
\bibitem{book1} R. Byron Baird, Warren E. Stewart, Edwin N. Lightfoot, "Transport Phenomena", John Wiley \& Sons INC.
\bibitem{book2} S. Sarkar, H. Satz and B. Sinha, "The Physics of the Quark-Gluon Plasma: Introductory Lectures", Lect. Notes Phys. 785, Springer (2010).

\bibitem{bk2ref52} P. Romatschke, U. Romatschke, 
%Viscosity Information from Relativistic Nuclear Collisions: How Perfect is the Fluid Observed at RHIC?, 
Phys. Rev. Lett., {\bf 99}, 172301 (2007).
%arXiv:0706.1522 [nucl-th]
\bibitem{bk2ref53} M. Luzum and P. Romatschke, 
%Conformal Relativistic Viscous Hydrodynamics: Applications to RHIC Results at 200 GeV, 
Phys. Rev. C, {\bf 78}, 034915 (2008). %arXiv:0804.4015 [nucl-th]
\bibitem{bk2ref54} H. Song and U.W. Heinz, 
%Extracting the QGP Viscosity from RHIC Data ? A Status Report from Viscous Hydrodynamics,
J. Phys. G, {\bf 36}, 064033 (2009).
%arXiv:0812.4274 [nucl-th] 361
\bibitem{bk2ref51} K. Dusling, D. Teaney, 
%Simulating Elliptic Flow with Viscous Hydrodynamics, 
Phys. Rev.  C, {\bf 77}, 034905 (2008).
%arXiv:0710.5932 [nucl-th] 361.
\bibitem{etalat} H. Meyer, Phys. Rev. D, {\bf 76}, 101701 (2007); arXiv:hep-lat/0704.1801 (2007).

\bibitem{Laine1} M. Laine, O. Philipsen, M. Tassler, P. Romatschke, J. High Energy Phys., {\bf 03}, 054 (2007).
\bibitem{Wolschin} F. Nendzig, G. Wolschin, Phys. Rev. C, {\bf 87}, 024911 (2013); arXiv:hep-ph 1210.8366v1 (2012).
\bibitem{vitev1} R. Sharma, I. Vitev, Phys. Rev. C, {\bf 87}, 044905 (2013); arXiv:hep-ph/1203.0329 (2013).
\bibitem{vitev2} Samuel Aronson et al. Physics Lett B, 778, 384-391 (2018).
\bibitem{Strickland1} M. Strickland, D. Yager-Elorriaga,  J. Comput. Phys., 229:6015, 6026 (2010); arXiv:quant-ph/0904.0939 (2010).
\bibitem{Strickland2} M. Strickland, Phys. Rev. Lett., 107:132301 (2011); arXiv:hep-ph/1106.2571 (2011).
\end{thebibliography}
\end{document}